\documentclass[twocolumn,resetfootnote]{aastex7}

\usepackage[version=4]{mhchem} 
\usepackage[T1]{fontenc}
\DeclareTextSymbolDefault{\dh}{T1}

\newcommand\solmass{$M_{\odot}$}

\newcommand{\poseidon}{\texttt{POSEIDON}}
\newcommand{\eureka}{\texttt{Eureka!}}
\newcommand{\exotic}{\texttt{ExoTiC-JEDI}}

\usepackage[inline]{enumitem}
\usepackage{amssymb}

\hypersetup{linkcolor=violet,citecolor=teal}

\begin{document}

\title{GEMS JWST: A sub-Solar metallicity atmosphere for giant planet TOI-5293Ab orbiting a rapidly changing M-dwarf}

\author[0000-0001-8401-4300]{Shubham Kanodia}
\affiliation{\Carnegie}
\email{skanodia@carnegiescience.edu}

\author[0000-0003-4835-0619]{Caleb I. Ca\~nas}
\altaffiliation{NASA Postdoctoral Program Fellow}
\affiliation{\Goddard}
\email{c.canas@nasa.gov}

\author[0000-0002-0746-1980]{Jacob Lustig-Yaeger}
\affiliation{\APL}
\email{jacob.lustig-yaeger@jhuapl.edu}

\author[0000-0001-6340-8220]{Giannina Guzm\'an Caloca}
\affiliation{\Goddard}
\affiliation{Department of Astronomy, University of Maryland, College Park, MD 20742, USA}
\email{gguzmanc@umd.edu}

\author[0000-0003-0354-0187]{Nicole L. Wallack}
\affil{\Carnegie}
\email{nwallack@carnegiescience.edu}

\author[0000-0002-8278-8377]{Simon M\"{u}ller}
\affiliation{\Zurich}
\email{simonandres.mueller@uzh.ch}

\author[0000-0001-5555-2652]{Ravit Helled}
\affiliation{\Zurich}
\email{rhelled@physik.uzh.ch}


\author[0000-0001-8020-7121]{Knicole D. Col\'on}
\affil{\Goddard}
\email{knicole.colon@nasa.gov}

\author[0000-0002-1483-8811]{Ian Czekala} 
\affiliation{Australia Telescope National Facility, CSIRO Space and Astronomy, PO Box 76, Epping, NSW 1710, Australia}
\email{ian.czekala@csiro.au}

\author[0000-0003-1439-2781]{Megan Delamer}
\affiliation{\PSUAA}
\affiliation{\PSUCEHW}
\email{mmd6393@psu.edu}

\author[0000-0002-7127-7643]{Te Han}
\affiliation{Department of Physics \& Astronomy, The University of California, Irvine, Irvine, CA 92697, USA}
\affiliation{Kavli Institute for Astrophysics and Space Research, Massachusetts Institute of Technology, Cambridge, MA 02139, USA}
\email{tehan@mit.edu}

\author[0000-0002-2990-7613]{Jessica Libby-Roberts}
\affiliation{\PSUAA}
\affiliation{\PSUCEHW}
\affiliation{Department of Physics and Astronomy, University of Tampa, Tampa, FL 33606, USA}
\email{jlibbyroberts@ut.edu}

\author[0000-0002-4487-5533]{Anjali A. A. Piette}
\affiliation{School of Physics \& Astronomy, University of Birmingham, Edgbaston, Birmingham B15 2TT, UK}
\email{a.a.a.piette@bham.ac.uk}

\author[0000-0002-7352-7941]{Kevin B. Stevenson}
\affiliation{\APL}
\email{Kevin.Stevenson@jhuapl.edu}

\author[0000-0001-7409-5688]{Gu\dh mundur Stef\'ansson}
\affil{Anton Pannekoek Institute for Astronomy, University of Amsterdam, Science Park 904, 1098 XH Amsterdam, The Netherlands}
\affiliation{Astrophysics and Space Institute, Schmidt Sciences, New York, NY 10011, USA}
\email{g.k.stefansson@uva.nl}

\author[0009-0008-2801-5040]{Johanna Teske} 
\affiliation{\Carnegie}
\affiliation{The Observatories of the Carnegie Institution for Science, 813 Santa Barbara St., Pasadena, CA 91101, USA}
\email{jteske@carnegiescience.edu}

\newcommand{\Goddard}{NASA Goddard Space Flight Center, 8800 Greenbelt Road, Greenbelt, MD 20771, USA}
\newcommand{\Carnegie}{Earth and Planets Laboratory, Carnegie Science, 5241 Broad Branch Road, NW, Washington, DC 20015, USA}
\newcommand{\Zurich}{Department of Astrophysics, University of Z\"urich, Winterthurerstrasse 190, 8057 Z\"urich, Switzerland}
\newcommand{\PSUAA}{Department of Astronomy \& Astrophysics, The Pennsylvania State University, 525 Davey Laboratory, University Park, PA 16802, USA}
\newcommand{\PSUCEHW}{Center for Exoplanets and Habitable Worlds, The Pennsylvania State University, 525 Davey Laboratory, University Park, PA 16802, USA}
\newcommand{\APL}{Johns Hopkins APL, 11100 Johns Hopkins Rd, Laurel, MD 20723, USA}


\correspondingauthor{Shubham Kanodia}
\email{skanodia@carnegiescience.edu}

\begin{abstract}
The growing sample of Giant Exoplanets around M-dwarf Stars (GEMS) helps probe the extremes of giant planet formation. Comparing the properties of this sample with their FGK counterparts can help us understand how planet formation and migration depend on stellar mass. We initiated a large Cycle 2 JWST transmission spectroscopy survey of seven GEMS. Here we present the atmospheric characterization using two JWST transits of TOI-5293Ab, a 0.5 $M_J$ planet orbiting an early M-dwarf with a period of $\sim$ 3 days. The two NIRSpec/PRISM transits indicate the planet is eclipsing a rapidly changing (heterogeneous) stellar photosphere. We see that Visit 1 had heterogeneity crossings across the entire transit chord, rendering inferences from it to be unreliable. The Visit 1 spectrum exhibits a downward slope ${<1}$ $\mu$m suggestive of stellar contamination from faculae. In contrast, for Visit 2 we are able to model the heterogeneity crossings and obtain a transmission spectrum free from stellar contamination. We therefore limit our conclusions to a detailed analysis of Visit 2, and using Bayesian free chemistry retrievals, we find a low atmospheric metallicity ($\log [\mathrm{M/H}] = -1.03^{+0.53}_{-0.44}$ $\times$ Solar) and slightly super-solar C/O ratio ($1.23^{+2.94}_{-0.75}$). The retrievals yield Bayes factors that indicate strong evidence for \ce{CH4} as well as low significance detections of \ce{CO2}, \ce{H2O}, \ce{NH3}. Finally, using thermal evolution models we find that the radius of TOI-5293Ab is inflated above theoretical expectations ($\sim$ 1.07 $R_J$), despite it having an temperature of $\sim$ 700 K, and hence we were unable to constrain its bulk composition.
\end{abstract}




\section{Introduction} 
In the quest to understand exoplanets --- and subsequently contextualize them against our Solar system planets --- transiting planets enable measurements of planetary masses and radii, and hence their bulk densities. However, comparing just bulk properties is prone to degeneracies in interior structure, composition, and more. The next frontier has been the detection and characterization of planetary atmospheres \citep{seager_theoretical_2000, charbonneau_detection_2002} through transmission spectroscopy. In particular, detections of molecular spectral signatures in planetary atmospheres can enable quantification of their abundances\footnote{See Appendix in \citet{fortney_framework_2013} for conversion between mass fractions and metallicities.} (e.g., C/H, O/H), their metallicity (M/H) and elemental ratios (e.g., C/O, C/N). These estimates can be contrasted with equilibrium predictions to ascertain the impact of disequilibrium processes like vertical mixing in the atmosphere (i.e., convection) and photochemistry, as well as the presence of aerosols. Abundance estimates can also collectively be compared across planet samples to test theoretical predictions across dimensions such as stellar host mass, equilibrium temperature, planet mass, etc., that we think influence the formation and evolution of exoplanets. 

One such dimension that has been relatively poorly studied (up until recently) is the influence of the host stellar mass on giant planet properties. NASA's TESS mission \citep{ricker_transiting_2014} has enabled dedicated searches for Giant Exoplanets around M-dwarf stars (GEMS) such as \citet{gan_occurrence_2023, triaud_m_2023, hartman_toi_2024, kanodia_searching_2024}. The properties of these objects have started to stretch our understanding of giant planet formation \citep[][etc.]{kanodia_toi-5205b_2023, hotnisky_searching_2025} -- they necessitate efficient accretion of rocky material in the early stages of giant planet formation to initiate runaway gaseous accretion in a timely manner before gaseous disk dissipation \citep{laughlin_core_2004, sanchez_formation_2026}. \cite{kanodia_transiting_2024} compared the bulk properties of these warm giant planets ($<$ 1000 K) across the stellar mass axis from 0.35 -- 1.2 \solmass{}, and found them to be agnostic of the host star mass, suggesting a commonality in the formation processes of giant planets across host star mass. However, as mentioned earlier, comparisons of bulk properties can be reductive and prone to degeneracies in composition and interior temperature. To remedy this, we utilized JWST \citep{gardner_james_2006, clampin_status_2008} to study the atmospheres of seven GEMS with transmission spectroscopy \citep{greene_characterizing_2016} as part of our large Cycle 2 program (GO 3171) --- \textit{Red Dwarfs and the Seven Giants: First Insights Into the Atmospheres of Giant Exoplanets around M-dwarf Stars} \citep{kanodia_red_2023, canas_gems_2025}.  The goal of this program is to compare the atmospheric and bulk metallicities of this sample across both planet mass (intra-sample), and stellar mass (by comparing with FGK giant planet observations from JWST). 

As part of our Cycle 2 survey, here we present inferences from two JWST observations of TOI-5293Ab \citep{canas_toi-3984_2023}. The system consists of a short period ($P\sim2.93$ days) gas giant planet with a radius of $\sim$ 1 $R_J$, mass of 0.5 $M_J$, orbiting a $\sim$ 0.5 \solmass{} early M-dwarf with a rotation period of $\sim$ 20 days \citep{canas_toi-3984_2023}. It also includes a widely separated mid M-dwarf (TOI-5293B) at a projected separation of 3.5'' or $\sim$ 580 AU. Additionally, Rossiter-McLaughlin \citep{rossiter_detection_1924, mclaughlin_results_1924} measurements with MAROON-X \citep{seifahrt_development_2016, seifahrt_-sky_2020-1} have revealed a well-aligned orbit for the planet \citep[][]{weisserman_aligned_2025}.

In Section \ref{sec:observation} we describe the observations and data reduction, followed by forward modelling in Section \ref{sec:picaso} and Bayesian atmospheric retrievals in Section \ref{sec:retrievals}. We discuss the implications of our results in Section \ref{sec:discussion}, before finally concluding in Section \ref{sec:conclusion}.

\section{Observations \& Data Reduction}\label{sec:observation} 
We obtained two transits of TOI-5293Ab with JWST using NIRSpec/PRISM in Bright Object Time Series (BOTS) mode \citep{birkmann_-flight_2022} on 2024 June 27 and 2024 July 9 (observation numbers 11 and 20, respectively). TOI-5293A was used directly for target acquisition and each observation used the NRSRAPID readout pattern with the SUB32 array. All observations consisted of 14528 integrations (total duration of 5.56 hrs) with 5 groups per integration (median observational cadence of 1.38 s). Each observation contained the complete transit ($\sim1.94$ hours) and included at least one hour of out-of-transit baseline. No observation saturated or exceeded $\gtrsim75\%$ the well depth ($\sim49152$ ADU), which has been suggested as a conservative limit for the onset of non-linearity during the ramp fitting stage \citep[e.g.,][]{carter_benchmark_2024}. 

In this work, we analyzed the uncalibrated data files (level 0 \texttt{*uncal.fits} data products) available on the Barbara A. Mikulski Archive for Space Telescopes (MAST). The raw files were processed using two pipelines, \texttt{ExoTiC-JEDI}\footnote{\url{https://github.com/Exo-TiC/ExoTiC-JEDI}} \citep{alderson_early_2022} and \texttt{Eureka!}\footnote{\url{https://github.com/kevin218/Eureka}} \citep{bell_eureka_2022}, following steps similar to those outlined in \cite{canas_gems_2025}, which are summarized below.

\subsection{\texttt{ExoTiC-JEDI} reduction}
The first two stages of \texttt{ExoTiC-JEDI} wrap the \texttt{jwst} pipeline (v1.18.0 with CRDSv12.1.5 and context \texttt{jwst\_1364.pmap}) and follow the default procedures in the \texttt{jwst} pipeline to perform detector-level and spectroscopic processing, with a few exceptions. In Stage 1 (\texttt{calwebb\_detector1}), we skipped the \texttt{jump} step, which has been shown to be ineffective for observations with a small number of groups \citep[e.g.,][]{Rustamkulov2023}. We also skipped the \texttt{superbias} step in \texttt{jwst} and the analogous \texttt{custom\_bias} step in \texttt{ExoTiC-JEDI} because all observations used the same detector. Similar to \cite{canas_gems_2025}, we masked column 125 from all analysis due to the presence of a trace-adjacent hot pixel. To remove correlated noise at the detector readout level \citep[e.g.,  ``$1/f$ noise'' ;][]{birkmann_-flight_2022, alderson_early_2022, Rustamkulov2023}, we skipped the \texttt{clean\_flicker} step and instead employed the custom ``de-stripping'' step from \texttt{ExoTiC-JEDI} to estimate the background value as the median of pixels more than 10 Full-Width-at-Half-Maximum (FWHM) from the trace center and subtract this value column-by-column at the group level. For Stage 2 (\texttt{calwebb\_spec2}), we skipped the \texttt{photom} and \texttt{flat\_field} step, as our analysis did not require flux-calibrated spectra. 

After Stage 2, we replaced saturated, dead, bad, low quantum efficiency, and no gain pixels with the median value of the 5 pixels surrounding it. We also replaced outliers identified (i) spatially with a rejection threshold of 5$\sigma$ deviation from the median using the median of the surrounding 20 columns for replacement and (ii) temporally with a rejection threshold of 10$\sigma$ using the median of that pixel in the surrounding 10 integrations for replacement. 

In Stage 3, we identified the trace position by fitting a Gaussian to each column and adopted the center and FWHM from a fit to the Gaussian parameters with a zero-order polynomial. The box aperture extended to three times the FWHM (0.724 pixels) of the spectral trace for a full aperture width of $\sim6$ pixels. We masked all pixels $\sim10$ FWHMs from the trace center and used the remaining pixels (top and bottom $\sim6$ rows) to correct for residual $1/f$ and background noise by subtracting the median of the region column-by-column. We used an optimal extraction algorithm \citep{horne_optimal_1986} to produce 1D spectra,\footnote{We ignore any additional thresholding for cosmic rays as \texttt{ExoTiC-JEDI} replaces outliers before this step.} which were realigned for positional shifts that were determined by cross-correlation.

\subsection{\texttt{Eureka!} reduction}
As with the \texttt{ExoTiC-JEDI} pipeline, the first two stages of \texttt{Eureka!} are wrappers of the \texttt{jwst} pipeline (v1.18.0 with CRDSv12.1.5 and context \texttt{jwst\_1364.pmap}) to reduce the uncalibrated level 0 files. In Stage 1, we followed the default steps in the \texttt{calwebb\_detector1} pipeline with the exception of the jump step, which we skipped as we used outlier rejection algorithms in later stages. Similar to the \texttt{ExoTiC-JEDI} reduction, we masked column 125 from all analysis due to the presence of a trace-adjacent hot pixel. We retained the default \texttt{clean\_flicker\_noise} step to correct $1/f$ noise. We also performed group-level background subtraction using the median of the top and bottom 7 rows to estimate the background in each column. In Stage 2, we followed the default steps in the \texttt{calwebb\_spec2} pipeline with the exception of the \texttt{photom} and \texttt{flat\_field} steps, which were skipped because our analysis did not require calibrated spectra.

In Stage 3, we generated 1D stellar spectra following an optimal spectral extraction algorithm \citep{horne_optimal_1986}. We adopted a spectral half-width of 2 pixels and a background half-width of 7 pixels. Our Stage 3 background region estimated the median background value using all pixels along the top and bottom rows that were greater than 7 pixels from the trace center. We created a normalized spatial profile using the median of all data frames after removing outliers greater than 10$\sigma$. We performed optimal extraction using an outlier threshold of 30$\sigma$ \citep[see step 7 of][]{horne_optimal_1986}.

\begin{figure}[bb]
\epsscale{1.15}
\plotone{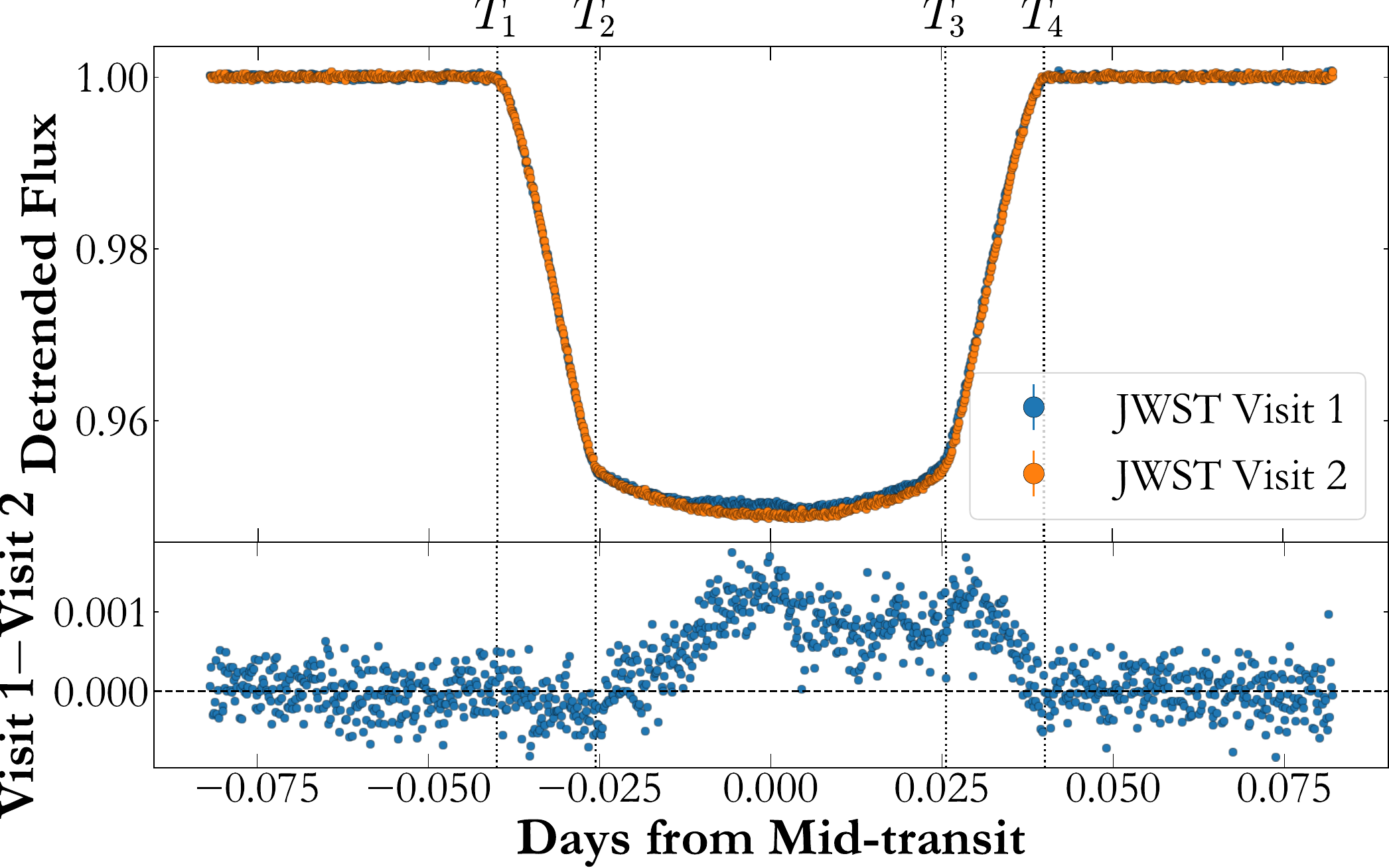}
\caption{The phase-folded white light curves from the \texttt{ExoTiC-JEDI} reduction, after detrending with a quadratic polynomial (see \autoref{tab:5293wlcpriors} for details) and binning to a cadence of 10 s for clarity. The labels $T_{1}-T_4$ mark the four separate points of contact between the planetary and stellar disks (e.g., start of transit, end of ingress, start of egress, end of transit). \textbf{Takeaway:} Visit 1 displays a different transit shape and depth (\autoref{apptab:NoSpotTransitfit}), suggesting that complex surface heterogeneities during Visit 1 impart signals on the \textit{entirety} of the transit, including subtle differences at the limbs.} 
\label{fig:visits_overplot}
\end{figure}

\subsection{Light curve generation}\label{app:wlcgen}
The white light curves for the \texttt{ExoTiC-JEDI} reduction are over-plotted and compared in \autoref{fig:visits_overplot}, where we note that Visit 1 differs starkly from Visit 2 (with a maximum difference of $\sim1300$ ppm between the light curves near mid-transit). In particular, Visit 1 contains heterogeneities across the entire transit chord starting from first contact ($T_1$) to fourth contact ($T_4$). We also note that the two visits are separated by $\sim11.7$ days or $\sim0.57\times P_{\mathrm{rot}}$ such that the stellar surface may be significantly different between each visit and contribute to the observed differences in transit shape.

\subsection{White light curve fits with the \texttt{ExoTiC-JEDI} reduction}\label{sec:wlcfits}

We performed a fit \textit{only} to the \texttt{ExoTiC-JEDI} white light curve. For other planets in this survey we have seen that we obtain comparable spectra when performing white light curve fits across different reductions (Wallack et al. in prep.), and most importantly, the retrieved abundances do not vary \citep[explored here in Section \ref{sec:retrievals:individual}; as well as for previous planets][]{canas_gems_2025}. For computational efficiency, these fits were performed on the white light curves after binning to a cadence of 5 s and only on data $\pm1.5$ transit durations from mid-transit ($\pm1.94$ hrs).

\subsubsection{Simple Transit Fit}\label{sec:batman}

\begin{figure}[tt]
\epsscale{1.15}
\plotone{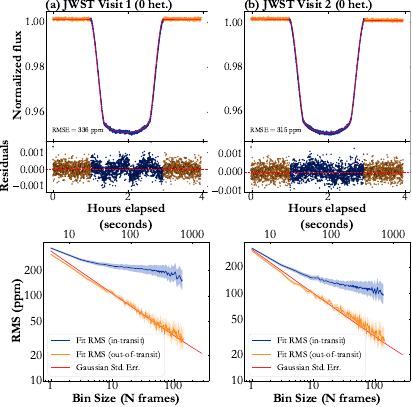}
\caption{JWST NIRSpec/PRISM white light curves produced using \texttt{ExoTiC-JEDI}. Data are binned to a cadence of 5 s. \textbf{Top row}: The data and best-fitting model (solid line) along with the residuals to the fit. In-transit data are blue while out of transit data are orange. \textit{These models do not account for surface heterogeneities}. \textbf{Bottom row}: The RMS for each Visit for in-transit (blue) and out-of-transit (orange) data. The prediction for Gaussian white noise is shown as a red solid line. The black point at (0, 0.001) represents the median error of the data. The analogous models including surface heterogeneities are displayed in \autoref{fig:transitcrossing}. \textbf{Takeaway:} Note the stark differences in residuals to a simple transit fit between the visits.}
\label{fig:NoSpotTransitFit}
\end{figure}

 We first fit a transit to the white light curve using \texttt{juliet} \citep{espinoza_juliet_2019} assuming no heterogeneities for Visits 1 and 2 (\autoref{fig:NoSpotTransitFit}) such that the light curves were modelled using \texttt{batman} \citep{kreidberg_batman_2015}. We adopted a quadratic limb darkening law in which the limb darkening coefficients were sampled using broad Gaussian priors ($\sigma=0.1$) that had a mean equal to the prediction from \texttt{ExoTiC-LD} \citep{exoticld} using the limb-angle dependent specific intensities of the \texttt{NewEra} stellar library \citep{hauschildt_newera_2025}. The free parameters for the white light curve fits were the orbital parameters ($P$, $T_0$, $a/R_\star$, $b$), the scaled planetary radius ($R_p/R_\star$), quadratic limb darkening coefficients ($q_1$, $q_2$) sampled following \cite{kipping_efficient_2013}, and the coefficients for the quadratic polynomial to model systematic trends ($c_0$, $c_1$, $c_2$). We used the orbital parameters from \cite{canas_toi-3984_2023} as priors for the white light curve fits. These fits and all subsequent fits with \texttt{juliet} (i) assumed no jitter or dilution terms in the models (dilution is fixed to unity and jitter is assumed to be zero) and (ii) used \texttt{dynesty} \citep{Speagle2020} to perform dynamic nested sampling with a convergence criterion of $\Delta\ln Z=0.01$. 
 
 While the results are summarized in \autoref{apptab:NoSpotTransitfit}, we note some salient points of discrepancy between the visits: (i) the ratio of planetary to stellar radii ($R_p/R_*$) is $\sim3\sigma$ lower in Visit 1 than for Visit 2, which is consistent with the residuals seen in \autoref{fig:visits_overplot}, (ii) the combination of spot crossings at the limbs and lower $R_p/R_*$ also bias the recovered impact parameter, $b$, and the limb darkening coefficients to be $>3\sigma$ discrepant between visits, suggesting that the overall transit shape was different between both visits.


 Due to the continuous heterogeneity crossings along almost the entire transit chord, we consider the Visit 1  white light curve fit to be unreliable due to the lack of robust photospheric (uncontaminated) baseline during the transit (between $T_2$ and $T_3$) or at the limbs. Given the issues with M-dwarf limb darkening parameters \citep[see appendix in     ][]{canas_gems_2025}, it is challenging to place theoretical constraints on the limb darkening (instead of fitting them). Furthermore, we see heterogeneity crossing events in Visit 2 as well --- though they do not cover the entire transit chord --- and hence we cannot trivially adopt orbital parameters from Visit 2 to Visit 1. \textbf{Out of abundance of caution and to not misinterpret biased data, while we present results and discussion for visit 1 and 2 for completeness and comparison, our final inferences and conclusions regarding the planetary atmosphere are limited to results from Visit 2.}

\startlongtable
\begin{deluxetable*}{lccccc}
\tabletypesize{\footnotesize}
\tablecaption{System parameters for TOI-5293A and priors for fits to the spectroscopic light curves assuming a circular orbit. In particular, note the stark contrast between Visits 1 and 2.\label{tab:5293wlcpriors}}
\tablehead{\colhead{Parameter} &
\colhead{Units} &
\multicolumn2c{Value} &
\colhead{Reference}
}
\startdata
\sidehead{Stellar parameters:}
~~~Stellar Mass ($M_\star$) & $\mathrm{M_\odot}$ & \multicolumn2c{$0.54 \pm 0.02$} & 1\\
~~~Stellar Radius ($R_\star$) & $\mathrm{R_\odot}$ & \multicolumn2c{$0.52_{-0.01}^{+0.02}$} & 1\\
~~~Effective Temperature ($T_{\mathrm{eff}}$) & K & \multicolumn2c{$3586\pm88$} & 1\\
~~~Surface Gravity ($\log g_\star$) & dex & \multicolumn2c{$4.77\pm0.05$} & 1\\
~~~Metallicity ($\mathrm{[Fe/H]}$) & dex & \multicolumn2c{$-0.03\pm0.12$} & 1\\
~~~Rotation period ($P_\mathrm{rot}$) & days & \multicolumn2c{$20.6_{-0.4}^{+0.3}$} & 1\\
\hline\hline
\noalign{\vskip 0.75ex} White light curve fits: & & Prior$^a$ & Visit 1 & Visit 2  \\
\noalign{\vskip 0.75ex}\hline
~~~Period & days & Fixed & \multicolumn2c{2.930289} \\
~~~Time of mid-transit ($T_0$) & $\mathrm{BJD_{TDB}}$ & $\mathcal{N}(2460489.17,0.01)$ & $2460489.16605 \pm 0.00002$ & $2460489.165974 \pm 0.000003$ \\
~~~Scaled semi-major axis ($a/R_\star$) & \nodata & $\mathcal{N}(14.1,2.0)$ & $13.95 \pm 0.02$ & $14.13 \pm 0.01$ \\
~~~Scaled radius ($R_p/R_\star$) & \nodata & $\mathcal{U}(0,1)$ & $0.2162_{-0.0007}^{+0.0006}$ & $0.2211_{-0.0004}^{+0.0003}$\\
~~~Impact parameter ($b$) & \nodata & $\mathcal{U}(0,1)$ & $0.183 \pm 0.008$ & $0.087_{-0.01}^{+0.008}$\\
~~~Inclination ($i$) & deg & \nodata & $89.25 \pm 0.03$ & $89.65_{-0.03}^{+0.04}$\\
~~~Linear limb darkening coefficient ($q_1$) & \nodata & $\mathcal{N}(0.115,0.1)$ & $0.25 \pm 0.01$ & $0.143_{-0.006}^{+0.007}$\\
~~~Quadratic limb darkening coefficient ($q_2$) & \nodata & $\mathcal{N}(0.136,0.1)$ & $0.03 \pm 0.01$ & $0.21 \pm 0.01$ \\
~~~Polynomial baseline constant term ($c_0$) & \nodata & $\mathcal{U}(-3,3)$ & $0.025 \pm 0.006$ & $0.053_{-0.003}^{+0.002}$ \\
~~~Polynomial baseline linear term ($c_1$) & \nodata & $\mathcal{U}(-3,3)$ & $-0.00013 \pm 0.00009$ & $-0.00477_{-0.00009}^{+0.0001}$ \\
~~~Polynomial baseline quadratic term ($c_2$) & \nodata & $\mathcal{U}(-3,3)$ & $0.014 \pm 0.004$ & $-0.002 \pm 0.004$ \\
\hline\hline
\noalign{\vskip 0.75ex} Free parameters in spectroscopic light curve fits: & & \multicolumn{3}{c}{Prior}  \\
\noalign{\vskip 0.75ex}\hline
~~~Scaled radius ($R_p/R_\star$) & \nodata & \multicolumn{3}{c}{$\mathcal{U}(0,1)$} \\
~~~Errorbar Scaling ($\sigma_{\mathrm{JWST}}$) & \nodata & \multicolumn{3}{c}{$\mathcal{U}(0,5)$} \\
~~~Linear limb darkening coefficient ($u_1$)$^b$ & \nodata & \multicolumn{3}{c}{$\mathcal{N}(\mathtt{NewEra},0.1)$} \\
~~~Quadratic limb darkening coefficient ($u_2$)$^b$ & \nodata & \multicolumn{3}{c}{$\mathcal{N}(\mathtt{NewEra},0.1)$} \\
~~~Polynomial baseline constant term ($c_0$) & \nodata & \multicolumn{3}{c}{$\mathcal{U}(-3,-3)$} \\
~~~Polynomial baseline linear term ($c_1$) & \nodata & \multicolumn{3}{c}{$\mathcal{U}(-3,-3)$} \\
~~~Polynomial baseline quadratic term ($c_2$) & \nodata & \multicolumn{3}{c}{$\mathcal{U}(-3,-3)$} \\
~~~Spot flux ratio ($f_\mathrm{spot}$) & \nodata & \multicolumn{3}{c}{$\mathcal{U}(0,2)$}\\
\enddata
\tablerefs{1) \cite{canas_toi-3984_2023}}
\tablenotetext{a}{$\mathcal{N}(X,Y)$ defines a Gaussian prior with a mean of $X$ and standard deviation of $Y$. \\ $\mathcal{U}(X,Y)$ defines a uniform prior between a lower limit of $X$ and upper limit of $Y$.}
\tablenotetext{b}{Quadratic limb darkening coefficients for the spectroscopic light curve fits using \texttt{Eureka!} were sampled as $u_1$ and $u_2$ centered on the values predicted with \texttt{ExoTiC-LD}.}
\end{deluxetable*}

\subsubsection{Heterogeneity + Transit Fit}\label{sec:spotfit}

\begin{figure*}[tt]
\epsscale{0.8}
\plotone{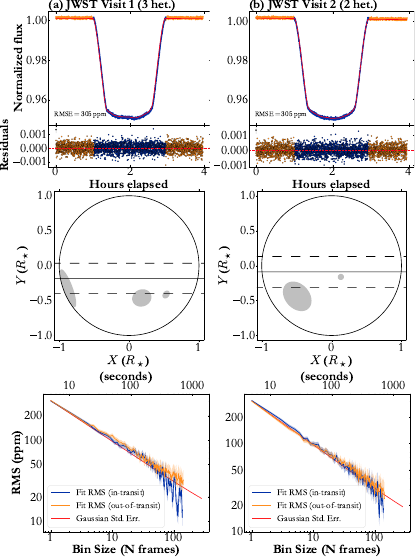}
\caption{Similar to \autoref{fig:NoSpotTransitFit} but using \texttt{spotrod} to model surface heterogeneities. \textbf{Top row}: The data and best-fitting model (solid line) along with the residuals to the fit. \textbf{Middle row}: The stellar surface and the adopted heterogeneity configuration. The transparency of the heterogeneities \textit{do not} reflect the flux ratio (we do not differentiate between spots or faculae). The solid line marks the center of the planet and the dashed lines mark the transit chord ($\pm~R_p$). \textbf{Bottom row}: The RMS for each visit for in-transit (blue) and out-of-transit (orange) data along with the median error of the data (black point).}
\label{fig:transitcrossing}
\end{figure*}

Next, we attempted to account for the visible stellar contamination in our data. We followed a similar procedure to \cite{canas_gems_2025} by using a modified version of \texttt{juliet} that implemented \texttt{spotrod} \citep{beky_spotrod_2014} to model the white light curves and determine an optimal number of stellar surface heterogeneities (spots or faculae, varying between $0-6$). Unlike the procedure in \cite{canas_gems_2025}, each visit was fit independently and each heterogeneity was allowed to have a unique flux ratio, $0\le f_\mathrm{het}$\footnote{The convention used by \texttt{spotrod} is that the photosphere has a flux ratio of unity while a feature is classified as a cool spot if $f_\mathrm{het}<1$ or hot facula if $f_\mathrm{het}>1$.}$\le2$.  

The free parameters for the surface heterogeneities were sampled using
\begin{enumerate*}[label=(\roman*)]
\item a uniform prior on the flux ratio ($0\le f_\mathrm{het}\le2$),
\item a uniform prior on the radius ($0\le r_\mathrm{het}\le0.5$),
\item and a central position ($X_{\mathrm{het}},Y_{\mathrm{het}}$) that was sampled from a unit disk.
\end{enumerate*} 
The remaining free parameters and priors are identical to the simplified white light curve fit described in Section \ref{sec:batman}.

Due to the complexity that each additional surface heterogeneity added to the model (an additional four free parameters) and the multimodal posteriors often observed when modelling these features \citep[e.g.,][]{canas_gems_2025,2025arXiv251103045M}, we considered model selection criteria (see Appendix \autoref{apptab:spotevidence}) that penalized more complex models with a large number of free parameters, including the AIC \citep[Akaike Information Criterion;][]{akaike_information_1973} and BIC \citep[Bayesian Information Criterion;][]{Schwarz1978}. We adopted 3 features in Visit 1 and 2 features in Visit 2  based on the BIC because this metric uniquely yielded strong evidence \citep[$\Delta \mathrm{BIC}>5$ compared to all other configurations;][]{Trotta2008} for only one configuration. The priors and posteriors for the fits to the white light curves are listed in \autoref{tab:5293wlcpriors} with the best-fitting configuration of the surface heterogeneities in \autoref{apptab:spotpar}. \autoref{fig:transitcrossing} presents the best-fitting white-light transit model.

\subsection{Transmission spectra}
We used \texttt{Eureka!} Stage 4 for both reductions to generate the spectroscopic light curves at (i) pixel-level resolution, with one light curve for each column spanning indices $43-451$ (0.53 \textmu{}m $\lesssim\lambda\lesssim$ 5.3 \textmu{}m), and (ii) a resolution binned to 4-pixels ($\sim$ 40 nm bin widths), which produced 118 channels that spanned 0.53 \textmu{}m $\lesssim\lambda\lesssim$ 5.3 \textmu{}m. For both binning schemes, we excluded outliers that were identified using a five-iteration $4\sigma$ rejection step applied to the photometric time series after smoothing with a rolling median filter (width of 10 integrations).

Similar to \citet{canas_gems_2025}, we fit the spectroscopic light curves from the \texttt{ExoTiC-JEDI} and \texttt{Eureka!} reductions using a modified version of \texttt{Eureka!} that implemented a fit with \texttt{spotrod} in Stage 5. For these fits, the parameters of the surface heterogeneities and orbital elements were fixed to the best-fitting values derived from the \texttt{ExoTiC-JEDI} white light curve. The following orbital parameters were fixed during the fits to the spectroscopic light curves: $P$, $T_0$, $e$, $\omega_\star$, $a/R_\star$, $i$, feature size and location (see \autoref{tab:5293wlcpriors} and \autoref{apptab:spotpar}). As with the white light curve fits (see Section \ref{sec:spotfit}), we adopted an identical Gaussian prior centered on the predictions from the PHOENIX  \texttt{NewEra} stellar spectra library \citep{hauschildt_newera_2025} for the quadratic limb darkening coefficients. We included an identical systematic baseline as the white light curve fits (a quadratic polynomial). The fits to the spectroscopic light curves had the following free parameters:
\begin{enumerate*}[label=(\roman*)]
\item radius ratio ($R_p/R_\star$),
\item quadratic limb darkening coefficients ($u_1,~u_2$),
\item coefficients for the baseline ($c_0$, $c_1$, $c_2$),
\item flux ratio for each heterogeneity ($f_\textrm{het}$), and
\item an error multiplicative factor ($\sigma_{\mathrm{JWST}}$).
\end{enumerate*}
We list the priors for these parameters in \autoref{tab:5293wlcpriors}. 
Similar to the white light curve fits, we performed dynamic nested sampling with \texttt{dynesty} with 1000 live points and a convergence criterion of $\Delta\ln Z=0.01$. 

\begin{figure}
    \centering
    \includegraphics[width=\linewidth]{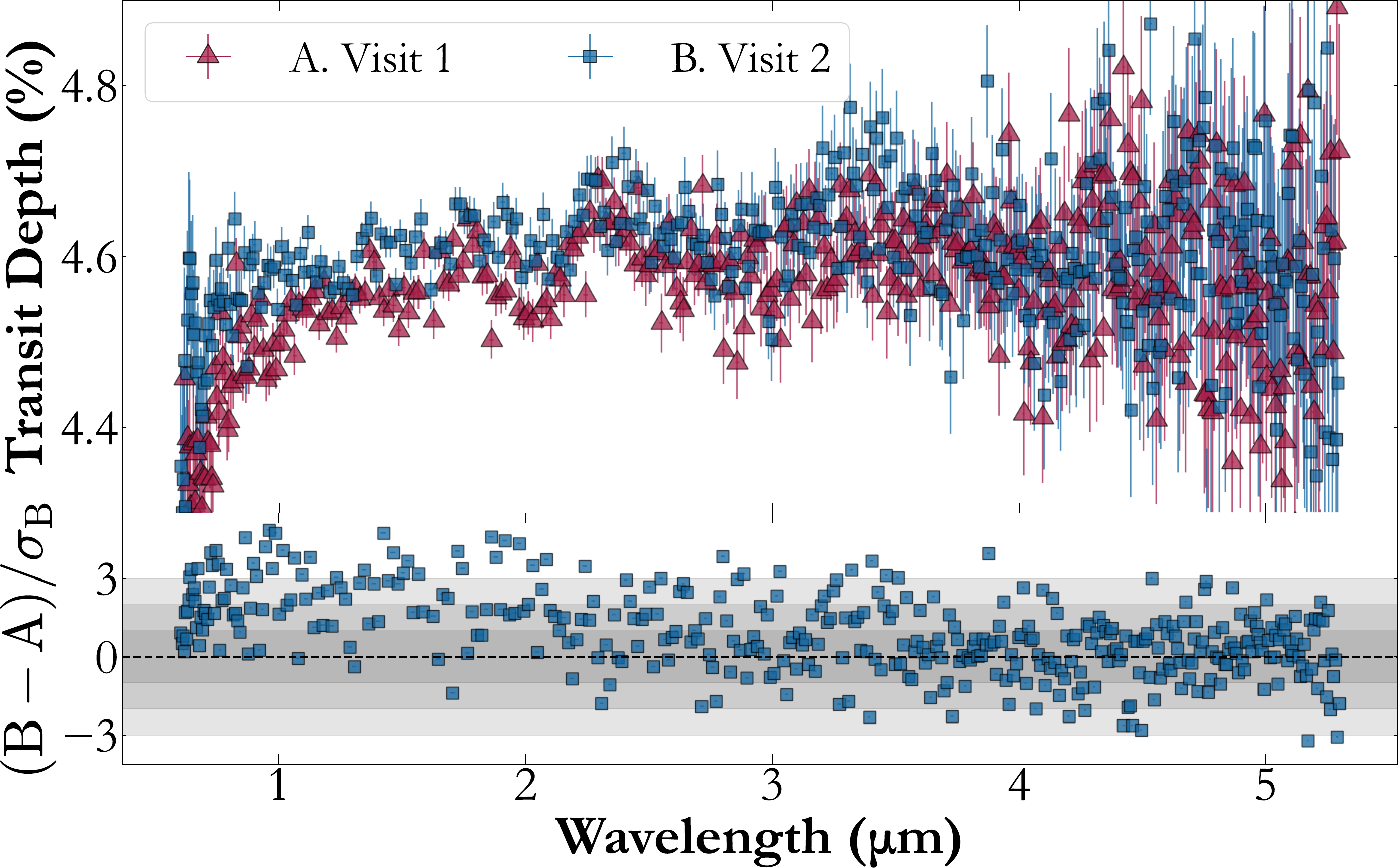}
    \caption{\textbf{Top:} The pixel-level \texttt{ExoTiC-JEDI} transmission spectra with Visit 1 as red triangles and Visit 2 as blue squares. \textbf{Bottom:} The differences between both visits, scaled by the errors of the first visit (median $\sigma_B\sim630$ ppm). The 1, 2, and 3$\sigma$ regions are shaded for reference. \textbf{Takeaway:} As mentioned earlier, we see substantial differences in the transit properties and transmission spectra between visits, especially at wwavelengths $\lesssim3.8$ \textmu{}m.}
    \label{fig:TransmissionSpectra}
\end{figure}

\begin{figure*}[!t]
    \centering
    \includegraphics[width=\linewidth]{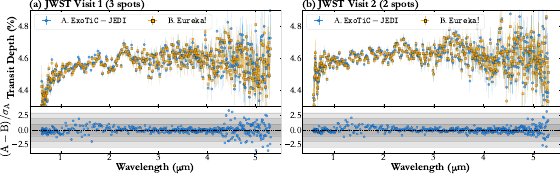}
    \caption{\textbf{(a)}. The pixel-level \texttt{ExoTiC-JEDI} transmission spectrum (blue circles) compared to the \texttt{Eureka!} pixel-level transmission spectrum (orange squares) for Visit 1. The difference is shown in the bottom row and scaled by the errors of the \texttt{ExoTiC-JEDI} data. \textbf{(b)}. The same but for Visit 2. }
    \label{fig:TransmissionSpectraComp}
\end{figure*}

The transmission spectra for both visits from \texttt{ExoTiC-JEDI} are shown in \autoref{fig:TransmissionSpectra} and \texttt{Eureka!} is compared to \texttt{ExoTiC-JEDI} in \autoref{fig:TransmissionSpectraComp}. Between Visits 1 and 2, the data display similar absorption features but with an observed difference (ranging between $1-3\sigma$) for $\lambda\lesssim3.8$ \textmu{}m.\footnote{The observed difference between visits 1 and 2 decreases as a function of wavelength. There is a median offset between the spectra of $\sim1000$ ppm and $\sim30$ ppm between $0.6-0.9$ \textmu{}m and $3.6-3.8$ \textmu{}m, respectively.} This chromatic offset is likely the result of the complex spot structure seen in Visit 1 (see Section \ref{sec:batman}), which produces a different transit shape (e.g., limb darkening, impact parameter, radius, etc.) and would impact our spectroscopic light curves (in which we fit the orbital elements to the best fits from the white light curve). Although we did not derive a configuration for the heterogeneities for the \texttt{Eureka!} reduction, the \texttt{ExoTiC-JEDI} and \texttt{Eureka!} spectra agree at $2\sigma$ for the NIRSpec/PRISM bandpass, suggesting that the underlying reductions produced nearly identical light curves. Given the agreement between the \texttt{ExoTiC-JEDI} and \texttt{Eureka!} reductions, the chromatic difference between the visits is most likely due to the evolution of the surface heterogeneities. For the remainder of this manuscript, we only consider the analysis on Visit 2, which has less contamination in the white light curve fits. Accounting for the rapid evolution of stellar variability and de-contamination of the first visit is beyond the scope of this manuscript. We provide a comparison of the limb darkening and spot contrasts for each visit from the \texttt{ExoTiC-JEDI} reduction in \autoref{app:ldspot}. Although we adopt \texttt{ExoTiC-JEDI} as the primary data reduction for analysis, we compare the \texttt{Eureka!} reductions for Bayesian retrievals in Section \ref{sec:visitretrieval}.


\section{Forward models assuming equilibrium chemistry}\label{sec:picaso}

We modelled the \texttt{ExoTiC-JEDI} transmission spectrum between $0.6$ \textmu{}m $\le\lambda\le$5.3 \textmu{}m using the \texttt{PICASO} package\footnote{\url{https://github.com/natashabatalha/picaso}} \citep[v3.3;][]{batalha_exoplanet_2019,Mukherjee2022}. The one dimensional climate solver implemented in \texttt{PICASO} relies on correlated-k opacities\footnote{\url{https://zenodo.org/records/7542068}} \citep[see][]{lupu_correlated_2021} derived with the Sonora Bobcat models \citep{marley_sonora_2021}. We used the \cite{Guillot2010} pressure-temperature profile as the initial guess to generate a grid of radiative-convective thermochemical equilibrium (RCTE) atmospheric models covering:
\begin{enumerate*}[label=(\roman*)]
\item 13 atmospheric metallicities ($\log \mathrm{[M/H]}= -1.0$, $-0.7$, $-0.5$, $-0.3$,~0.0,~0.3,~0.5,~0.7,~1.0,~1.3,~1.5,~1.7,~2.0 dex),
\item 6 carbon-to-oxygen ratios ($\mathrm{C/O}=0.25$, 0.5, 1.0, 1.5, 2.0, 2.5 relative to solar $\mathrm{[C/O]_\odot=0.458}$),
\item 5 intrinsic temperatures ($T_{\mathrm{int}}=50$, 100, 200, 300, 400 K), and
\item 3 heat redistribution factors ($r_{st}=0.25,~0.50,~0.75$).
\end{enumerate*}
The nominal grid would contain 1170 models, however, only 1140 models converged.\footnote{The failed models consisted of points where both $\mathrm{[M/H]}\ge1.5$ dex and $\mathrm{C/O}\le0.5$ relative to solar.} We used \texttt{PICASO} to generate transmission spectra from these profiles with an opacity database sampled to $R=60,000$ \citep{picasodb2020} from the original $R\sim10^6$ line-by-line calculations presented in \cite{Freedman2008} and \texttt{EXOPLINES} \citep{exoplines2021}.

\begin{figure*}[]
\centering
\includegraphics[width=1.0\linewidth] {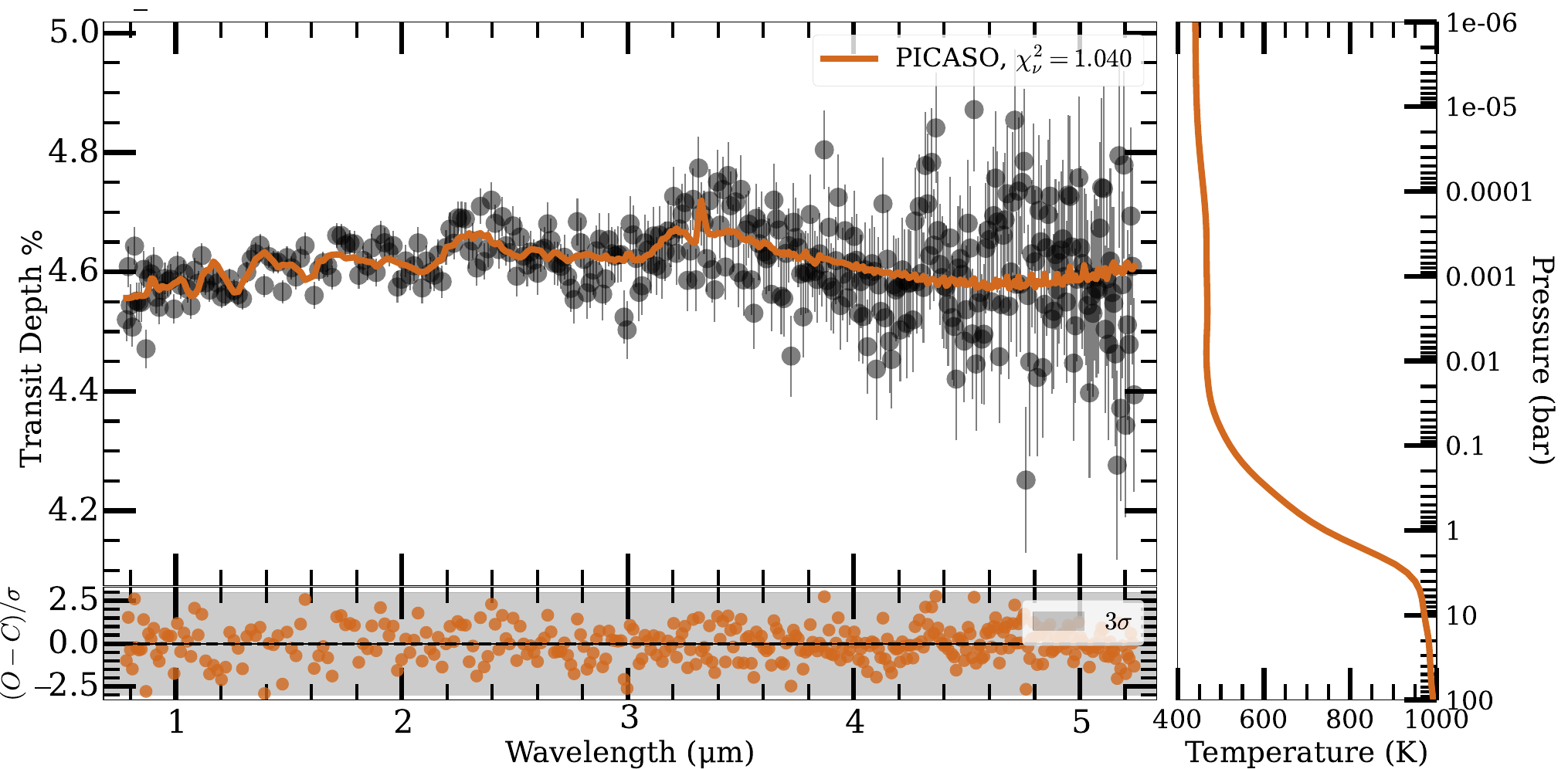} 
\caption{\textbf{Top Left.} The TOI-5293Ab pixel-level transmission spectrum for Visit 2 derived with \texttt{ExoTiC-JEDI}, plotted with the best-fitting grid-based model assuming equilibrium chemistry (orange line, obtained from \texttt{PICASO}). \textbf{Bottom Left.} The difference between the data and model, scaled by the errors of the data. The $\pm3\sigma$ region is shaded for reference. \textbf{Right.} The pressure-temperature profile associated with the best-fitting model. \textbf{Takeaway:} We obtain a good equilibrium chemistry fit from \texttt{PICASO} for the unbinned \texttt{ExoTiC-JEDI} pixel-level spectrum that finds a metallicity that reaches the lower limit of the bounds $\mathrm{[M/H]}\to-1.0$ dex Solar and upper limits of carbon-to-oxygen $\mathrm{C/O}\to2.5$ $\times$ Solar. }
\label{fig:picasovulcan}
\end{figure*}

We employed the $\chi^2_\nu$ grid search method implemented in \texttt{PICASO} to fit the Visit 2 spectrum. The best-fitting model from the RCTE grid had $\mathrm{[M/H]}=-1.0$ dex Solar, $\mathrm{C/O}=2.5$ relative to solar\footnote{The C/O reported by the Sonora Bobcat models \citep{marley_sonora_2021} is a multiple of $\mathrm{[C/O]}_\odot=0.458$ based on the protosolar abundances from \cite{Lodders2010} and, for the best-fitting model, is equivalent to $\mathrm{C/O}=1.145$ or $\log\mathrm{[C/O]}=0.059$ dex.}, $T_{\mathrm{int}}=50$ K, and $r_{st}=0.25$. This fit reached the limits of the metallicity ($\mathrm{[M/H]}\to-1.0$) and carbon-to-oxygen ($\mathrm{C/O}\to2.5$) parameter space covered by the correlated k-tables \citep{lupu_correlated_2021} that were used, such that we could not exclude an even lower atmospheric metallicity or higher C/O. Although we did not explore stellar contamination or aerosol opacity with these model grids, we note that the RCTE grid yielded $\chi^2_\nu=1.04$, suggesting that the data could be well-described by a clear atmosphere in thermochemical equilibrium. Qualitatively, the low atmospheric metallicity and super-solar carbon-to-oxygen ratio are consistent with the results from other planets in the GEMS survey \citep[e.g.,][]{canas_gems_2025}. To explore aerosol opacities, the impact of stellar contamination, and to ensure no biases due to the limits of the \texttt{PICASO} model grids, we also performed a set of Bayesian atmospheric retrievals (next section).

\section{Atmospheric Retrievals}\label{sec:retrievals}
We utilize Bayesian atmospheric retrievals to understand the transmission spectrum of TOI-5293Ab and quantify the statistical significance of the features present therein \citep{madhusudhan_atmospheric_2018}. Based on transmission spectra of other planets in this survey \citep[Guzm\'an Caloca et al., submitted; Ashtari et al., submitted;][]{canas_gems_2025}, as well as results from the \texttt{PICASO} grid retrieval in \autoref{fig:picasovulcan}, we opt to directly run free chemistry retrievals. This is because the [M/H] and C/O seen here are beyond the range probed by equilibrium chemistry grid in the \poseidon{} \footnote{https://github.com/MartianColonist/POSEIDON}\citep{macdonald_hd_2017, macdonald_poseidon_2023} generated using\texttt{FastChem} \citep{stock_fastchem_2018}, which spans a range of -1 $\le$ log[M/H] $\le$ 4 and 0.2 $\le$ C/O $\le$ 2.  Given the similarity between the \texttt{ExoTiC-JEDI} and \eureka{} spectra for the two transits, we elect to run all our retrieval tests on the pixel-level (unbinned) \exotic{} spectra. We perform all the retrievals with opacities sampled to $R=10,000$ in the wavelength range of 0.7 -- 5.2 $\mu$m and include the following species: \ce{H2O}, \ce{CH4}, \ce{CO2}, \ce{CO}, \ce{SO2}, \ce{H2S}, \ce{NH3}, \ce{HCN} and \ce{C2H2} with line lists from ExoMol\footnote{\citet{polyansky_exomol_2018}, \citet{yurchenko_exomol_2024}, \citet{yurchenko_exomol_2020}, \citet{li_rovibrational_2015}, \citet{underwood_exomol_2016}, \citet{azzam_exomol_2016}, \citet{coles_exomol_2019}, \citet{barber_exomol_2014}, \citet{chubb_exomol_2020}.}, as well as collision-induced absorption (CIA) from HITRAN \citep{gordon_hitran2024_2026} for \ce{H2-H2}, \ce{H2-He}, \ce{H2-CH4}, \ce{CO2-H2}, \ce{CO2-CO2}, and \ce{CO2-CH4}. Based on the limited pressure range probed by transmission spectra, we use an isothermal pressure-temperature profile and isochemical profile for the molecules. This choice is further substantiated by the mostly isothermal pressure-temperature profile recovered from the \texttt{PICASO} grid retrieval in the 0.1 to 0.001 bar regime (\autoref{fig:picasovulcan}). 

The retrieval models that include a stellar contamination component to correct for stellar heterogeneities, i.e., the difference in stellar spectra between the occulted region (transit chord) and unocculted regions, follow the transit light source (TLS) effect framework from \citet{rackham_transit_2018}. To do this, we use the PHOENIX  \texttt{NewEra} stellar spectra library \citep{hauschildt_newera_2025} and the interpolation scheme from \texttt{pysynphot}\footnote{https://pysynphot.readthedocs.io/en/} \citep{stsci_development_team_pysynphot_2013}. Our TLS models include a stellar photosphere fit by its effective temperature ($T_{phot}$) and surface gravity ($\log g_{phot}$) along with a two component heterogeneity model that fits for the presence of spots and faculae with coverage fraction, $f_{\mathrm{spot}}$ and $f_{\mathrm{fac}}$ respectively, and assume that the remaining stellar area ($1-f_{\mathrm{spot}}-f_{\mathrm{fac}}$) is the photosphere\footnote{Here it is important to note that under this standard TLS framework, the fraction of spots and faculae is assumed to be the same in the occulted and unocculted region, which might not necessarily be the case and should be explored in future studies.}. The spot and faculae components are modelled assuming stellar spectra of effective temperature ($T_{fac}$ $>$ $T_{phot}$ and $T_{spot}$ $<$ $T_{phot}$), and surface gravity ($\log g_{spot}$ and $\log g_{fac}$) respectively, for a total of 8 additional free parameters. Given the complex instrumental and astrophysical systematics that are potentially not being accounted for in the measurement errorbars, we also include an error inflation term $b$ that is included following Equation 3 from \citet{line_uniform_2015}.

For models with the planetary atmosphere, we also include \textit{potential} sources of attenuation with clouds and hazes \citep{macdonald_hd_2017}. The clouds are modelled with a grey deck (achromatic) at a cloud-top pressure ($\rm \log P_{cloud}$), whereas the power law haze is characterized in terms of the Rayleigh enhancement factor relative to \ce{H2} Rayleigh scattering ($\log \rm a$) and the haze power law exponent ($\gamma$).

We sample our parameters using the nested sampling routine \texttt{MultiNest} \citep{feroz_multinest_2009}, via \texttt{PyMultiNest} \citep{buchner_x-ray_2014}. We use 1000 live points with the default convergence criterion of $\Delta\ln Z=0.5$. Additionally we evaluate model performance and test for the statistical significance using Bayesian evidence ($\mathcal{Z}$), and compare model performance for the nested models using Bayes factors ($\mathcal{B}$). While the model comparison and selection is performed solely on the basis of the Bayesian evidence, for intuitive comparison we also report an equivalent frequentist significance \citep{sellke_calibration_2001} in \autoref{tab:retrieval_models}, while advising caution regarding its interpretation \citep{thorngren_bayesian_2025}.

\begin{deluxetable*}{c|l|l|l|l}
\tablewidth{80pt}
\tabletypesize{\small}
\tablecaption{Model selection  \label{tab:retrieval_models}}
\tablehead{
\colhead{Name} & \colhead{Model Description} & \colhead{$\ln Z$} & \colhead{$\ln \mathcal{B}$} & \colhead{Significance}
}
\startdata
\multicolumn{5}{l}{Visit 1}  \\ \hline
M1 & TLS only (2 spots, free $\log(g)$) & 2190.9  &  22.2 &  7.0$\sigma$ \\ 
\hline
M2.1 & Atm only &  2210.2 & 2.9 & 2.9$\sigma$ \\
M2.2 & Atm + Cloud &  2210.2 & 2.9 & 2.9$\sigma$  \\
M2.3 & Atm + Cloud + Haze &  2209.5 & 3.6 & 3.2$\sigma$ \\
\hline
\textbf{M3.1} & \textbf{TLS$^\dagger$ + Atm }& \textbf{2213.1} & 0.0 & --- \\
M3.2 & TLS$^\dagger$ + Cloud Deck & 2217.2 & -4.1 & --- \\
M3.3 & TLS$^\dagger$ + Cloud Deck + Haze &  2217.2  & -4.1 & ---  \\
\hline
\hline
\multicolumn{5}{l}{Visit 2}  \\ \hline
M1 & TLS only (2 spots, free $\log(g)$) &  2180.1  & 28.1 & 7.8$\sigma$ \\ 
\hline
\textbf{M2.1} & \textbf{Atm only} & \textbf{2208.2} & 0.0 & ---   \\
M2.2 & Atm + Cloud & 2208.0 & 0.2 & 1.4$\sigma$    \\
M2.3 & Atm + Cloud + Haze & 2207.5 & 0.7 &  1.8$\sigma$ \\
\hline
M3.1 & TLS$^\dagger$ + Atm &  2205.8  & 1.4 & 2.7$\sigma$ \\
M3.2 & TLS$^\dagger$ + Cloud Deck & 2206.5 & 1.7 & 2.4$\sigma$  \\
M3.3 & TLS$^\dagger$ + Cloud Deck + Haze & 2206.8 & 1.4  & 2.3$\sigma$  \\
\enddata 
\tablenotetext{\dagger}{Model The stellar contamination model for the M3.X series includes a photosphere with two heterogeneities, each with free temperatures and surface gravities such that M3.X is a combination of models M1 and M2.X.}
\tablecomments{ The adopted model for Visit 1 is M3.1, and M2.1 for Visit 2, are marked in bold. The significances and $\ln \mathcal{B}$ for each model are with respect to the adopted model in bold.}
\end{deluxetable*}

We perform three tiers of retrievals, where (i) M1 considers no planetary atmosphere, but assigns the entire observed transmission spectra to stellar contamination from the TLS effect, (ii) M2.x considers planetary atmosphere models without any stellar contamination, and (iii) M3.x refers to models that include both a planetary atmosphere and stellar contamination (\autoref{tab:retrieval_models}). 
While we include free retrieval results for Visit 1 below for completeness, based on the aforementioned discussed complexities in modelling the heterogeneity crossing events (particularly in Visit 1), we perform independent model comparisons on the two visits (versus trying to combine them into one spectrum).

\subsection{Atmospheric retrievals on individual visits} \label{sec:retrievals:individual}\label{sec:visitretrieval}

As discussed in the previous section, Visit 1 displays heterogeneity crossing events throughout the transit (\autoref{fig:NoSpotTransitFit}), which further complicates the analysis, and thus we trust Visit 2 to be more reliable and discuss it first.

\begin{figure*}
    \centering
    \includegraphics[width=0.8\textwidth]{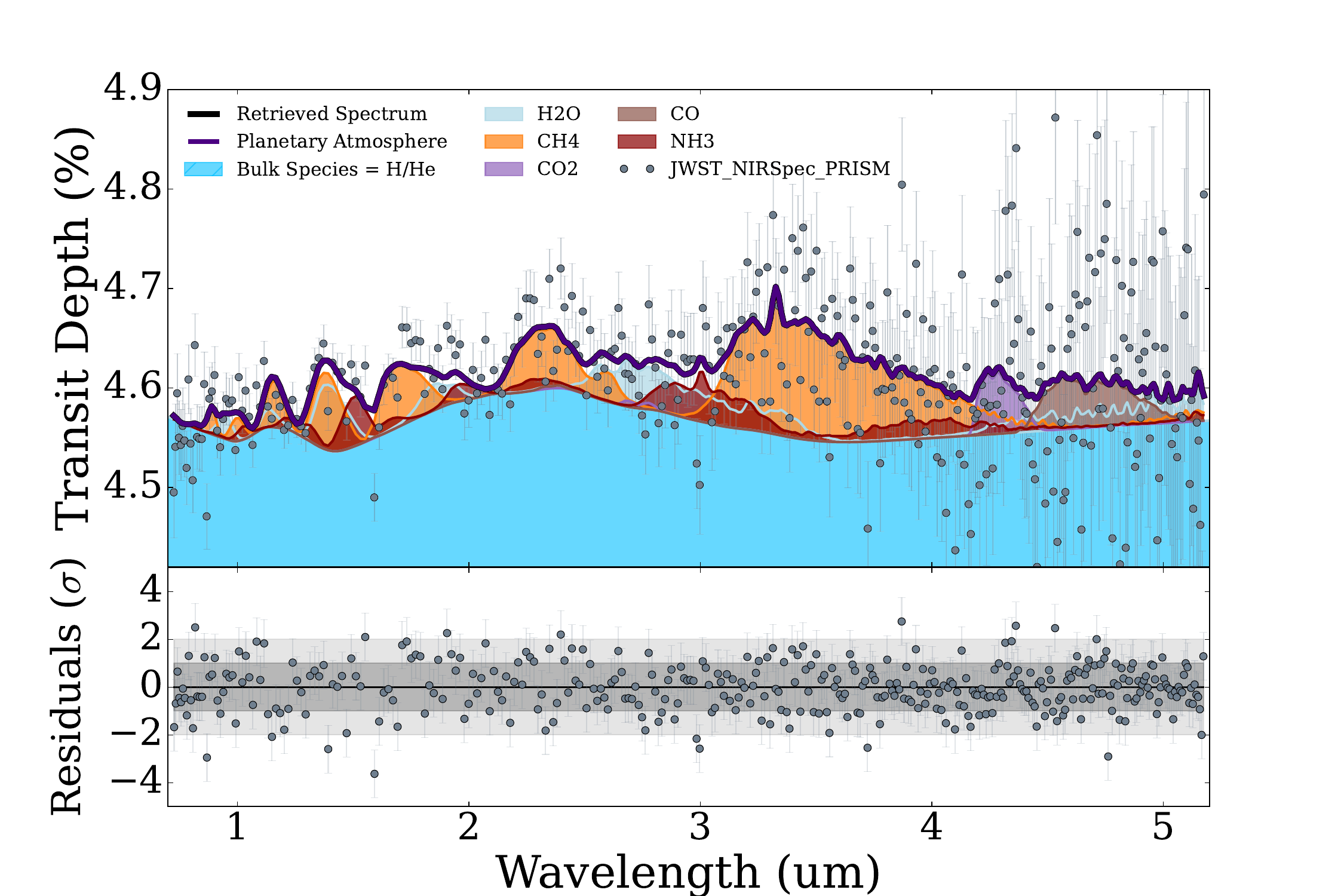}
    \hfill
    \includegraphics[width=0.5\textwidth]{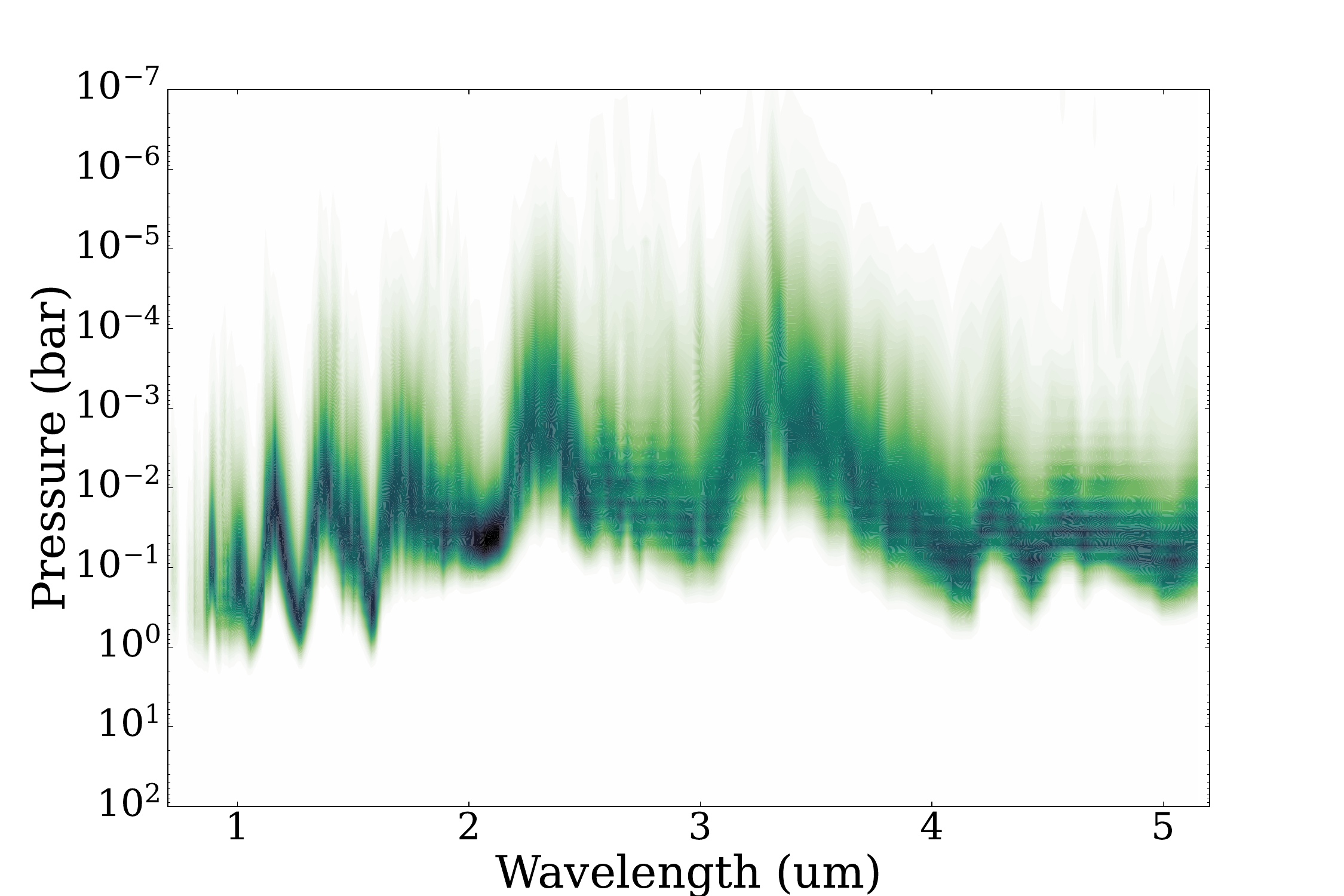}
    \caption{Retrieval results from the preferred models for Visit 2. \textbf{Top:} Maximum \textit{a posteriori} retrieved transmission spectrum (black line) with the contributions, derived using \texttt{POSEIDON}, shown from stellar contamination (red line) and atmospheric opacity (purple line). Alongside, contributions from individual species are shown in different colours (see legend). \textbf{Bottom:} Pressure contribution plot, where the darker regions signify pressures where the contribution from opacity sources in the transmission spectra is at its highest.
    \textbf{Takeaway:} While we note an unambiguous strong detection of \ce{CH4}, we also see hints of \ce{H2O}, \ce{NH3}, \ce{CO2}, though these face potential degeneracies that we discuss further.}
    \label{fig:retrieval_contributions_Visit2}
\end{figure*}

\textit{Visit 2:} Based on the log evidence, our most preferred model is M2.1 (\autoref{tab:retrieval_models}) indicating a spectrum dominated by the planetary atmosphere, as compared to a planetary atmosphere with significant stellar contamination (M3.x) or models with any attenuation from clouds or hazes (M2.2, M2.3). The M2.1 model has a $\chi^2_\nu = 1.05$, which is also similar to the \texttt{PICASO} grid fits. When compared with M1, we detect the clear evidence $> 7\sigma$ for the planetary atmosphere. We note that the model with clouds (M2.2) exhibits cloud-top pressure constraints that place them relatively deep in the atmosphere so that the interpretation compared to the cloud-free retrievals is effectively the same. The spectral fit (along with the contributions of individual species) is shown in \autoref{fig:retrieval_contributions_Visit2}. It is particularly interesting to note here that despite the clear heterogeneity crossings seen in the white light transit for Visit 2 (\autoref{fig:NoSpotTransitFit}), the retrievals do \textit{not} favour the inclusion of a TLS component, which suggests that there are not significant unocculted heterogeneities outside of the transit chord. This result is also in contrast to the preferred model in Visit 1 (M3.1). 

\begin{figure*}
    \centering
    \includegraphics[width=0.99\linewidth]{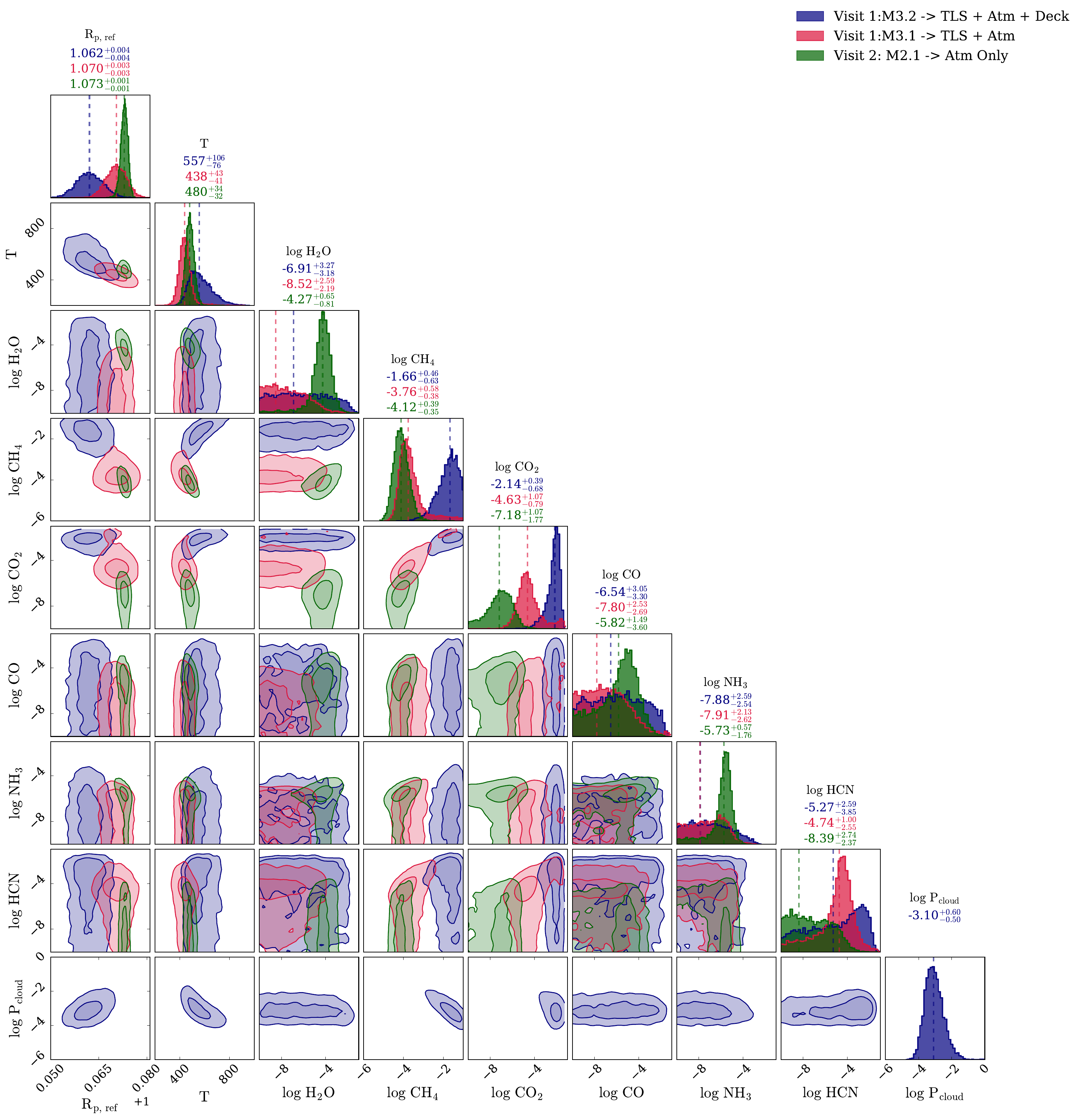}
    \caption{Given the complexities of spot fitting, results for Visit 2 are considered more reliable than Visit 1. For Visit 1,  M3.2 gives unphysical solutions (similarly high abundances for \ce{CH4} and \ce{CO2}) and degeneracies with temperature and cloud pressure, and is hence not chosen over M3.1. Furthermore, M3.1 for Visit 1 returns posteriors that are largely consistent with Visit 2, barring the retrieved abundance of \ce{H2O} that likely faces degeneracies with the TLS modelling.}
    \label{fig:CornerPlotComparison}
\end{figure*}

The retrieval results suggest the presence of \ce{CH4}, and potentially \ce{H2O}, \ce{CO2}, \ce{CO} and \ce{NH3} (full corner plot can be found in the Appendix \autoref{fig:CornerPlotVisit2_M2.1}, and we compare posteriors from the Visits in \autoref{fig:CornerPlotComparison}). We quantify the abundances and detection significance for these species in the next section. We note that the presence of water in the transmission spectra is degenerate with the lack of a TLS component (disfavoured at $\sim$2$\sigma$), and hence susceptible to inaccuracies in M-dwarf stellar spectra. This has been seen in previous GEMS JWST observations \citep{canas_gems_2025} as well as rocky M-dwarf planets \citep{moran_high_2023}.

We also repeated the final retrieval for Visit 2 (M2.1) using the 120-channel \eureka{} reduction and recover posteriors derived from the \exotic{} reduction for all the parameters consistent with our final recovered retrieval parameters at $<$ 1-$\sigma$ (listed in \autoref{tab:posteriors}).

\textit{Visit 1:} We see a clear preference for the transmission spectrum of Visit 1 to include stellar contamination and the planetary atmosphere (M3.x), in addition to which we note some complexities in the interpretation of the model comparison (\autoref{tab:retrieval_models}). The stellar contamination model fits the downward blue slope, which is indicative of the presence of heterogeneities hotter than the stellar photosphere, i.e., faculae, outside of the transit chord. While the statistically favoured model is M3.2, with a $\ln \mathcal{B} \sim$ 4 ($\sim$ 3$\sigma$) that includes a grey cloud deck, we note that this exacerbates a known degeneracy between the planet temperature and \ce{CH4} abundances (Appendix \autoref{fig:CornerPlotVisit1_M3.2}), which involves the scale height as high mean molecular weight gases transition between the trace gas and bulk gas regimes \citep[e.g., see][their Figure 6]
{novais_parameter_2025}. This degeneracy is further aggravated by the so-called ``normalization degeneracy'' described by \citet{heng_theory_2017} and \citet{fisher_retrieval_2018}, which stems from the unknown continuum offset and involves several parameters including the reference radius, cloud-top pressure, temperature, and now also, offsets from TLS modelling. This solution (M3.2) results in unphysically high abundances for \ce{CH4} and \ce{CO2} that are discrepant from the retrieved abundances for the more reliable Visit 2 (\autoref{fig:CornerPlotComparison}), in which clouds were not favoured at high altitude. We also note that the Atm-only (no TLS) solution for Visit 1 (M2.1) recovers a very high metallicity and \ce{CH4} abundance, which we interpret as providing clues to the response of the atmospheric retrieval to missing stellar physics from the favoured TLS model.  In other words, we consider that the high metallicities recovered in M3.2 for Visit 1 are likely due to deficiencies in the TLS models coupled to known retrieval degeneracies that are exacerbated in the presence of TLS. Given the consistency in recovered planetary parameters between M3.1 for Visit 1 and the preferred Visit 2 model (M2.1), we adopt M3.1 as the preferred model for Visit 1 (Appendix \autoref{appfig:retrieval_contributions_Visit1}), with posteriors shown in Appendix \autoref{fig:CornerPlotVisit1_M3.1} and repeating the caveats regarding spot complexities inhibiting robust and accurate estimation of transmission spectra for Visit 1.

We ignore any coadded spectra from the two visits given the differences seen in \autoref{fig:TransmissionSpectra} and we perform all subsequent analysis on the individual visits.

\subsection{Detection Significance} \label{sec:retrievals:abundances}

We perform model comparison analyses to estimate the detection significance of individual species for the preferred models in Visit 1 (M3.1) and Visit 2 (M2.1), where we re-run free chemistry retrievals, but exclude the species for which we are estimating the significance. Comparing the Bayes factors --- $\mathcal{B}$ --- between models, we estimate the odds that favour the presence of individual molecules (\autoref{tab:abundance_significance}).

\begin{deluxetable*}{l | cc | cc}
\tablecaption{Detection significance for molecular species estimated using systematic model comparison tests. Following recommendations from \cite{thorngren_bayesian_2025}, we consider $\ln \mathcal{B}$ $\ge$ 3.0 to be `Minor Results', $\ln \mathcal{B}$ $\ge$ 5.9 to be `Detections', and $\ln \mathcal{B}$ $\ge$ 14.4 as `Major Discoveries'.  \label{tab:abundance_significance}}
\tablehead{
\colhead{Species} &
\multicolumn2c{Visit 1 (M3.1)} &
\multicolumn2c{\textbf{Visit 2$^a$ (M2.1)}} \\
\cline{2-5}
\colhead{} & \colhead{$\ln Z$} & \colhead{$\ln \mathcal{B}$}
& \colhead{$\ln Z$} & \colhead{$\ln \mathcal{B}$}
}
\startdata
All Species & 2213.1 & --- & 2208.2 & --- \\
\ce{H2O} & 2210.6 & 2.5 & 2203.8 & 4.4 \\
\ce{CH4} & 2184.1 & 29  & 2168.5 & 39.7 \\
\ce{CO2} & 2207.0 & 6.1  & 2205.0 & 3.2 \\
\ce{CO} &  2210.4  & 2.7 & 2205.5 & 2.7 \\
\ce{NH3} & 2210.7  & 2.4 & 2204.7 & 3.5 \\
\ce{HCN} & 2210.5 & 2.6 & 2206.2 & 2.0  \\
\enddata
\tablenotetext{a}{Preferred Visit and model.}
\end{deluxetable*}

We see that even though the corner plots (\autoref{fig:CornerPlotVisit2_M2.1} and \autoref{fig:CornerPlotVisit1_M3.1}) suggest the presence of multiple species such as \ce{H2O}, \ce{CH4}, etc., in terms of the $\ln\mathcal{B}$, the nested models suggest a `Major Discovery' only for \ce{CH4} in both visits  \citep[adopting thresholds from ][]{thorngren_bayesian_2025}. We note a difference in the \ce{CO2} abundance between the two visits, which could be due to unresolved systematics in the preferred TLS models for Visit 1 (note the degeneracies between \ce{CO2} abundance and TLS parameters in \autoref{fig:CornerPlotVisit1_M3.1}). While we see a detection for \ce{CO2} in Visit 1 we do not consider it robust given the degeneracies discussed in the previous section. For Visit 2, we report a minor detection of \ce{H2O}, \ce{CO2}, and \ce{NH3}. For the atmospheric properties for TOI-5293Ab, we subsequently adopt the Visit 2 parameters from \autoref{tab:posteriors}.

\begin{deluxetable*}{r|l|l||c|c}
\tablewidth{0.98\textwidth}
\tablecaption{Atmospheric retrieval priors and posteriors ($\pm$ 1-$\sigma$, 16-84 percentile) for fit to the \texttt{ExoTiC-JEDI} reduction for Visit 1 (M3.1) and Visit 2 (M2.1) using \texttt{POSEDION}. The abundances are the volume mixing ratios for each molecular species, and upper limits reported are $<2\sigma$ (95th percentile).\label{tab:posteriors}}
\tablehead{
\colhead{Parameters} & \colhead{Units} & \colhead{Priors} & \colhead{Visit 1} & \colhead{\textbf{Visit 2$^a$}}}
\startdata
$\mathrm{R}_{\mathrm{p, \, ref}}$ & $\mathrm{R_J}$ & $\mathcal{U}(0.90, 1.22)$ & $1.070\pm0.003$ & $1.073\pm0.001$ \\
$\mathrm{T}$ & K & $\mathcal{U}(200.00, 1500.00)$ &  438$^{+43}_{-42}$ &  480$^{+34}_{-33}$ \\
$\log \, \mathrm{H_2 O}$ & dex & $\mathcal{U}(-12.00, -0.10)$ & -8.5$^{+2.6}_{-2.2}$ & -4.3$^{+0.7}_{-0.8}$ \\
$\log \, \mathrm{CH_4}$ & dex & $\mathcal{U}(-12.00, -0.10)$ & -3.8$^{+0.6}_{-0.4}$ & $-4.1\pm0.4$ \\
$\log \, \mathrm{CO_2}$ & dex & $\mathcal{U}(-12.00, -0.10)$ & -4.6$^{+1.1}_{-0.8}$ & -7.2$^{+1.1}_{-1.8}$ \\
$\log \, \mathrm{CO}$ & dex & $\mathcal{U}(-12.00, -0.10)$ & -7.8$^{+2.6}_{-2.7}$ & -5.8$^{+1.5}_{-3.6}$ \\
$\log \, \mathrm{SO_2}$ & dex & $\mathcal{U}(-12.00, -0.10)$ & -9.0$\pm1.9$ & -9.3$^{+1.9}_{-1.8}$ \\
$\log \, \mathrm{NH_3}$ & dex & $\mathcal{U}(-12.00, -0.10)$ & -7.9$^{+2.1}_{-2.6}$ & -5.7$^{+0.6}_{-1.8}$ \\
$\log \, \mathrm{H_2 S}$ & dex & $\mathcal{U}(-12.00, -0.10)$ & -6.5$^{+2.4}_{-3.4}$ & -8.6$^{+2.5}_{-2.3}$ \\
$\log \, \mathrm{HCN}$ & dex & $\mathcal{U}(-12.00, -0.10)$ & -4.7$^{+1.0}_{-2.5}$ & -8.4$^{+2.8}_{-2.4}$ \\
$\log \, \mathrm{C_2 H_2}$ & dex & $\mathcal{U}(-12.00, -0.10)$ & -7.4$^{+2.3}_{-2.9}$ & -9.2$^{+2.1}_{-1.9}$ \\
$\mathrm{f}_{\mathrm{spot}}$ & \nodata & $\mathcal{U}(0.00, 0.50)$ & 0.08$^{+0.07}_{-0.05}$ & \nodata \\
$\mathrm{f}_{\mathrm{fac}}$ & \nodata & $\mathcal{U}(0.00, 0.50)$ & 0.07$^{+0.06}_{-0.04}$ & \nodata  \\
$\mathrm{T}_{\mathrm{spot}}$ & K & $\mathcal{U}(3000, 3586)$ & 3201$^{+150}_{-117}$ & \nodata  \\
$\mathrm{T}_{\mathrm{fac}}$ & K & $\mathcal{U}(3586, 4800)$ & 3814$^{+159}_{-126}$ & \nodata  \\
$\mathrm{T}_{\mathrm{phot}}$ & K & $\mathcal{N}(3586, 176)$ & 3480$^{+86}_{-81}$ & \nodata  \\
$\log \, \mathrm{g}_{\mathrm{spot}}$ & cgs & $\mathcal{U}(3.50, 6.00)$ & 5.38$^{+0.35}_{-0.52}$ & \nodata  \\
$\log \, \mathrm{g}_{\mathrm{fac}}$ & cgs & $\mathcal{U}(3.50, 6.00)$ & 5.39$^{+0.37}_{-0.50}$ & \nodata  \\
$\log \, \mathrm{g}_{\mathrm{phot}}$ & cgs & $\mathcal{U}(3.50, 6.00)$ & 4.80$^{+0.33}_{-0.46}$ & \nodata  \\
$\mathrm{b}$ & \nodata &  $\mathcal{U}(-10.5, -3.5)$ & -7.56$^{+0.15}_{-0.17}$  & -7.73$^{+0.21}_{-0.36}$  \\ \hline
$\log~\mathrm{C/O}$$^b$ & dex & \nodata & $0.65^{+0.53}_{-0.47}$ & $0.09^{+0.53}_{-0.41}$ \\
$\log~[\mathrm{M/H}]$$^c$ & dex & \nodata & $-0.65^{+0.82}_{-0.47}$ & $-1.03^{+0.53}_{-0.44}$ \\
$\log~\mathrm{C/H}$ & dex & \nodata & $-0.25^{+0.74}_{-0.43}$ & $-0.73^{+0.47}_{-0.38}$ \\
$\log~\mathrm{O/H}$ & dex & \nodata & $-1.20^{+1.10}_{-0.75}$ &  $-1.09^{+0.66}_{-0.67}$ \\
$\log~\mathrm{N/H}$ & dex & \nodata & $-0.75^{+0.96}_{-1.35}$ & $-1.60^{+0.61}_{-0.72}$ \\
$\log~\mathrm{S/H}$ & dex & \nodata & $-1.55^{+2.18}_{-2.38}$ & $-3.08^{+1.69}_{-1.81}$ \\
\hline
\enddata 
\tablenotetext{a}{Preferred Visit and model.}
\tablenotetext{b}{Equivalent to [C/O] = $1.23^{+2.94}_{-0.75}$, compared to [C/O]$_{\odot}$ = $0.56^{+1.35}_{-0.34}$}
\tablenotetext{b}{Equivalent to [M/H] = $0.093_{-0.059}^{+0.222}$ $\times$ [M/H]$_{\odot}$ = $0.0010^{+0.0024}_{-0.0006}$}
\tablecomments{The radius is reported at a pressure level of 10 bar. The atmospheric metallicity ([M/H]) and elemental abundances (C/H, O/H, and S/H) are reported with respect to the solar values from \cite{lodders_solar_2020}. The adopted solar values are $[\mathrm{M/H}]_\odot=Z_\odot=0.011$, $\log~[\mathrm{C/H}]_\odot=-3.53$, $\log~[\mathrm{O/H}]_\odot=-3.27$, $\log~[\mathrm{N/H}]_\odot=-4.15$, and $\log~[\mathrm{S/H}]_\odot=-4.85$.}
\end{deluxetable*}

\section{Discussion}\label{sec:discussion}

\subsection{Verifying the Stellar Radius}
TOI-5293A is part of a binary system with an M-dwarf companion (TOI-5293B; Gaia DR3 2640121482094497024) at an angular separation of 3.57'' or projected separation of $\sim$ 580 AU \citep{canas_toi-3984_2023}. The close on-sky position of this companion could potentially contaminate the photometric magnitudes used in the stellar parameter estimation --- \cite{canas_toi-3984_2023} use APASS $BV$ \citep{henden_apass_2018}, Pan-STARRS1 $grizy$  \citep{chambers_pan-starrs1_2016}, 2MASS $JHK$ \citep{cutri_vizier_2003} and WISE $W123$ \citep{wright_wide-field_2010}. Given that the two stars are resolved in Gaia DR3 \citep{gaia_collaboration_gaia_2023}, \cite{weisserman_aligned_2025} attempted to account for the contamination by estimating the corrected $K_S$ magnitude based on the Gaia $G$mag and $B_P - R_P$ colour. However, not only can the Gaia $B_P$ estimate be fraught for such faint M-dwarfs, but also the background companion can potentially contaminate the prism spectra used to obtain these colour estimates. \cite{creevey_gaia_2022} mention that over different epochs a quasi-randomly oriented window 3.5'' $\times$ 2.1'' is used to extract the spectra that can have varying amounts of contamination, thereby potentially rendering $B_P - R_P$ estimates unreliable.

To validate this, we repeat the exercise from \cite{canas_toi-3984_2023} and fit a spectral energy distribution using \texttt{EXOFASTv2} \citep{eastman_exofastv2_2019} with only PSF extracted photometry from Pan-STARRS1, where the companion is spatially resolved and hence should not contaminate the photometry. We obtain a stellar mass and radius identical to that from \cite{canas_toi-3984_2023}. 

\begin{figure}
    \centering
    \includegraphics[width=0.9\linewidth]{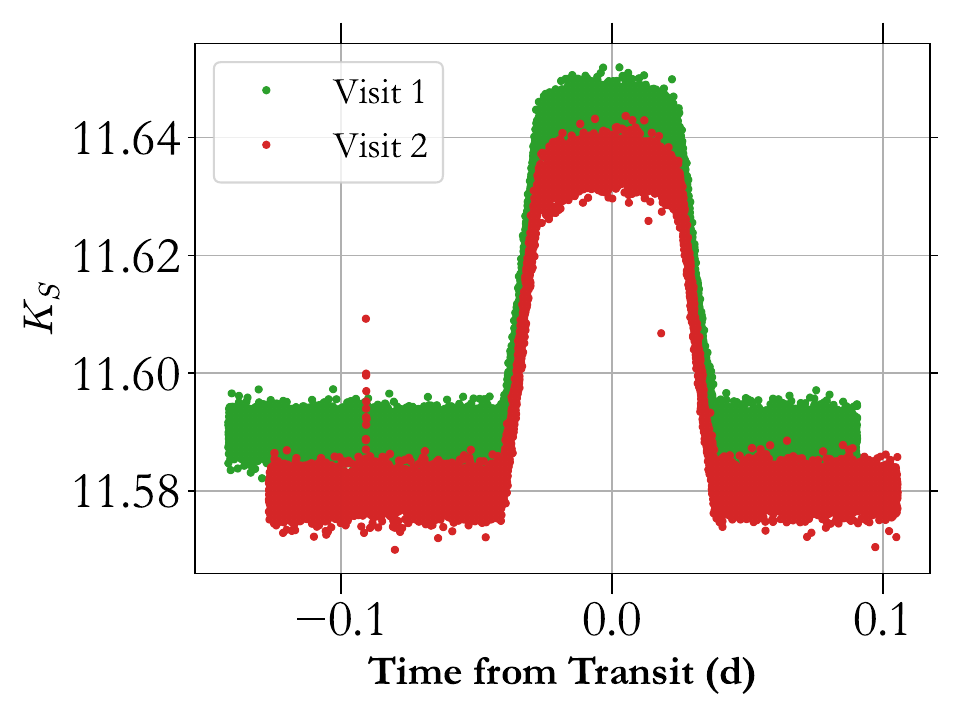}
    \caption{$K_S$ magnitude time-series for Visits 1 and 2 derived from NIRSpec/PRISM flux calibrated spectra by integrating over the filter bandpass. This derived $K_S$ estimate is uncontaminated by the neighbouring companion, and has a typical statistical error of 0.011 mag. It is consistent with the $K_S$ estimate used by \cite{canas_toi-3984_2023} of 11.64 $\pm$ 0.04.}
    \label{fig:FluxCalibratedKmag}
\end{figure}

Next, we use the stellar spectra obtained here using the NIRSpec BOTS 1.6'' $\times$ 1.6'' aperture \citep{birkmann_near-infrared_2022}, which spatially resolves and excludes TOI-5293B. Thus, we also perform a spectral flux calibration and obtain a time-series calibrated stellar spectra for Visits 1 and 2. To this end, we generate flux-calibrated stellar spectra using \texttt{Eureka!} in a similar manner as previously described when reducing the observations to determine the transmission spectra with a few notable differences. When extracting the spectra we use an aperture half-width of 8 pixels and a background region of 10 to ensure that we are fully capturing the entire stellar PSF. We also turn on the photon step within Stage 2 to determine the timeseries flux calibrated stellar spectra. We run Stage 1--3 with these changes and utilize the optimally extracted stellar spectra and associated errors from \texttt{Eureka!}. 
We then integrate these stellar spectra over the 2MASS $K_S$ bandpass to estimate the uncontaminated ${K_S}$ and associated uncertainties, and find the derived ${K_S}$ to be well within uncertainty between visits (\autoref{fig:FluxCalibratedKmag}). Importantly, our JWST derived ${K_S}$ magnitudes match the $K_S$ estimate from \cite{canas_toi-3984_2023} of 11.64 $\pm$ 0.04, and is brighter than `contamination corrected' 11.83 $\pm$ 0.04 used by \cite{weisserman_aligned_2025} to predict the stellar properties for TOI-5293A. Based on these two independent checks, we consider the stellar radius and mass estimates from \cite{canas_toi-3984_2023} to be reliable and use them throughout this work to constrain planetary properties.

\subsection{Atmospheric Inferences}
We limit our interpretation of atmospheric properties to Visit 2, with our final Bayesian retrieval posteriors listed in \autoref{tab:posteriors}. Evinced by \autoref{fig:retrieval_contributions_Visit2}, we see that the pressure levels probed by different sources of opacity across wavelength is largely in the $\sim$ $10^{-2}$ -- $10^{-1}$ bar range, which is deeper than pressures typically probed in transmission spectroscopy $\sim$ $10^{-3}$ bar and above \citep{rustamkulov_early_2022, schleich_knobs_2024}. This is likely due to the low-metallicity atmosphere, [M/H] $\sim$ -1 dex Solar (or 0.1\% absolute mass fraction), where the low opacity allows probing deeper in pressure at the terminators. 

The temperature recovered from the Bayesian retrievals is $\sim$500 K, which is much lower than the 675$^{+42}_{-30}$~K estimated by \cite{canas_toi-3984_2023} based on the stellar insolation and planetary semi-major axis. This difference could be due to a non-zero albedo, but is also similar to the grid retrieval from \texttt{PICASO} in \autoref{fig:picasovulcan} that suggests a temperature $\sim$ 500 K at the pressures probed by the transmission spectra (\autoref{fig:retrieval_contributions_Visit2}).

Next, we note that similar to previously published GEMS from this survey \citep[Guzm\'an Caloca et al., submitted; Ashtari et al., submitted;][]{canas_gems_2025}, TOI-5293Ab also seems to show a low metallicity, high C/O atmosphere --- log C/O $\sim$ 0.09, or [C/O] = $1.23^{+2.94}_{-0.75}$. Interestingly, despite the preferred model for Visit 2 not including a TLS component, we recover a very low abundance for \ce{H2O} (-4.3 dex VMR), and a corresponding low O/H of -1 dex compared to O/H$_{\odot}$, which suggests that the low \ce{H2O} abundance \textit{should} not be due to misattribution of the \ce{H2O} feature to the TLS effect \citep[e.g., as suggested as a possibility for earlier planets, e.g.,][]{moran_high_2023, canas_gems_2025}. Lastly, while we report minor detections of \ce{H2O} and also potentially \ce{NH3} based on their respective Bayes Factors, the most significant molecular detection is for \ce{CH4} that still returns C/H =  $-0.73^{+0.47}_{-0.38}$, that is sub-solar.

\subsection{Planetary interior}\label{bulkmetallicity}

The standard approach to characterize the interiors of exoplanets combines thermal evolution models with observational data \citep[e.g.,][]{fortney_interior_2010,thorngren_mass-metallicity_2016,muller_warm_2023}. Commonly thermal evolution models use the observed planetary mass and equilibrium temperature combined with varying bulk compositions to find the values that yield a matching planetary radius and age. In this work we used the \texttt{GASTLI} \citep{acuna_gastli_2024} open-source thermal evolution models to calculate the cooling of TOI-5293Ab to try and constrain its interior. In addition to the planetary mass, \texttt{GASTLI} requires the equilibrium temperature ($T_{\rm{eq}}$), atmospheric metallicity (log[M/H]), C/O ratio, and the core mass fraction. 

Since the atmospheric temperature of TOI-5293Ab at low pressures is significantly smaller than the zero-albedo equilibrium temperature of $T_{\rm{eq}} = 675$ K, we instead used the retrieved temperature of $\simeq 500$ K (see Figs. \ref{fig:picasovulcan} and \ref{fig:CornerPlotComparison}) as an estimate for the true equilibrium temperature. For the atmospheric metallicity we used the mean value of log[M/H] = -1.03, and for the C/O ratio we used 0.55. The final parameter to calculate a cooling curve with \texttt{GASTLI} is the core mass fraction. For the following calculation, we used a core mass fraction of zero to establish an upper bound for radius as a function of time.

\begin{figure}
    \centering
    \includegraphics[width=\linewidth]{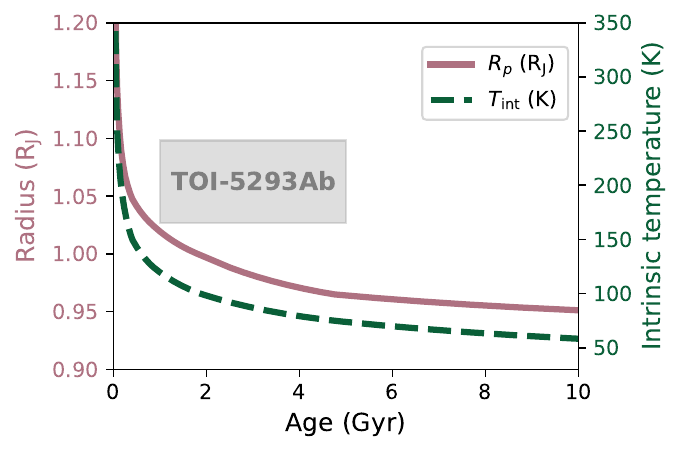}
    \caption{Radius (mauve) and intrinsic temperature (green) as a function of planetary age. The grey box shows the observational constraints for the radius and age of TOI-5293Ab. \textbf{Takeaway:} The calculated radius from the evolution model yields a smaller radius than observed at the system age of 1 to 5 Gyr, suggesting an anomalously inflated radii for TOI-5293Ab.}
    \label{fig:thermal_evolution}
\end{figure}

The radius and intrinsic temperature evolution for these parameters is shown in \autoref{fig:thermal_evolution}. It is clear that the radius of TOI-5293Ab is inflated, since the evolution model cannot match its observed radius despite allowing for a large uncertainty in the stellar age of 1 to 5 Gyr. Consequently we were not able to constrain the planetary bulk composition. We note that a previous study used the \texttt{planetsynth} \citep{muller_synthetic_2021} evolution models to infer either a zero or very low bulk metallicity for TOI-5293Ab \citep{muller_bulk_2025}. However, the statistical Monte Carlo approach used to derive the estimate likely relied on draws of the planetary radius that were more than one sigma away from the observed mean. 

These results suggest that TOI-5293Ab may be a part of the group of warm Jupiters that are inflated beyond what standard cooling models predict \citep{muller_theoretical_2020,kanodia_searching_2025,peerani_how_2026}. This requires either a mechanism to delay the cooling or an additional heating source \citep[e.g.,][]{leconte_is_2010,muller_warm_2023}. Interestingly, the results from the forward models (assuming equilibrium chemistry; see Section \ref{sec:picaso}) derived a low intrinsic temperature of about 50 K, which is roughly consistent with the cooling curve presented in \autoref{fig:thermal_evolution}. In this case, the radius inflation likely cannot be caused by an external energy source, since that should result in a higher intrinsic temperature \citep[e.g.,][]{thorngren_connecting_2019, sarkis_evidence_2021}. We also note here that this is not akin to the anomalous radius inflation seen for hot Jupiters, which takes place at much higher equilibrium temperatures \citep[$\sim$ 1000 K;][]{thorngren_bayesian_2018}.

The combined inflated radius and low intrinsic temperature could be explained by the presence of one or more thermal boundary layers trapping heat inside the planet. This would keep it inflated and result in a small intrinsic cooling luminosity. One common explanation for the inflation of planets is that composition gradients in the interior could prevent convection and therefore slowing the energy transport. However, we note that TOI-5293Ab appears to be inflated even when compared to an evolution model assuming a metal-free hydrogen-helium interior. To prevent convection with a composition gradient, heavy elements would have to be added to the planet, which increases its mean density and reduces the radius. It is therefore unlikely that the radius can be increased by adding heavy elements, even if the heat transport is impacted. A more promising mechanism that can significantly slow down the cooling of giant planets is a local increase in the opacity, for example if clouds are present \citep{poser_effect_2019,poser_effect_2024}, which is expected for warm giant planets orbiting M-dwarf stars \citep{kiefer_under_2024}, but not detected here (\autoref{tab:retrieval_models}).

\subsection{Suggestions for future observations}

It is particularly interesting that the transit shapes, transmission spectra, and retrieved parameters for Visit 1 and 2 (separated by $\sim$ 0.57 stellar rotation period) differ to the extent discussed above.  This suggests a sharp contrast in the spot/faculae distribution on the surface of TOI-5293Ab across its two hemispheres (as probed by successive visits). Therefore, we  recommend that for future observations of planets around M-dwarf stars, it would improve the robustness of stellar activity mitigation techniques to target transits observed at as similar stellar rotation phase as possible. Systems like TOI-5293A that have a known stellar rotation period from photometric modulation induced by heterogeneities \citep{canas_toi-3984_2023} already imply the presence of detectable surface heterogeneities (even in broadband photometry) that are further amplified for the TLS effect, particularly with the sensitivity of JWST. These lessons are critical for observations of small rocky planets around M-dwarf stars that typically have lower S/N, which can make disambiguating stellar and planetary signals particularly difficult.

\section{Conclusion}\label{sec:conclusion}
We present results from two JWST transits for the 0.5 $M_J$ giant planet orbiting TOI-5293Ab as part of the large Cycle 2 GEMS JWST survey. We note a change in the transit shape between the two visits caused by different heterogeneity crossing events that differ due to the temporal gap between visits being roughly half the stellar rotation period. This implies a remarkable difference in the photospheric properties of this host star across its two hemispheres. We consider inferences from Visit 1 to be unreliable due to the limitations in M-dwarf theoretical models that inhibit the accurate determination of spot and limb spectra that is required to obtain unbiased atmospheric properties. Fortunately, the second visit permits spot modelling and displays lesser stellar contamination. 

By forward modelling this spectrum, we find an atmosphere that seems to be close to equilibrium, with metallicity [M/H] approaching the \texttt{PICASO} lower bound of -1.0 dex Solar, C/O approaching its upper bound of 2.5 $\times$ solar or C/O = 1.145 and a low interior temperature. Bayesian free retrievals suggest that these observations probe deep into the planet atmosphere (10 -- 100 mbar), and find a comparable low sub-solar metallicity $-1.03^{+0.53}_{-0.44}$ (to Solar) and high [C/O] ($1.23^{+2.94}_{-0.75}$). These atmospheric chemistry inferences are driven by a significant detection of methane with a VMR of $-4.1 \pm 0.4$ dex ($\ln \mathcal{B} = 39.7$), and minor detections of water, ammonia, and carbon dioxide with $\ln \mathcal{B}$ of 4.4, 3.5, and 3.2 respectively. These are comparable to results for other planets from this survey and will enable a final quantitive intra-sample comparison.

Finally we note that beyond the scientific inferences for GEMS, these observations of giant exoplanets around M-dwarf stars, enabled by their large scale heights and transit depths, highlight the limitations of existing stellar contamination and limb darkening models. By spatially resolving spot and faculae crossing events as the planet crosses the heterogeneous transit chord, we are also able to independently validate results from the free-retrievals. We will compile these as recommendations in a future survey paper for M-dwarf planet host observations, particularly for the small rocky planets that offer the best opportunity for JWST to detect biosignatures.

\begin{acknowledgments}
We thank Peter Hauschildt for generating the \texttt{NewEra } center-to-limb variation spectra for M dwarfs.

This work is based on observations made with the NASA/ESA/CSA James Webb Space Telescope. The data were obtained from MAST at STScI, which is operated by the Association of Universities for Research in Astronomy, Inc., under NASA contract NAS 5-03127 for JWST. These observations are associated with program \#3171. Support for program \#3171 was provided by NASA through a grant from the Space Telescope Science Institute, which is operated by the Association of Universities for Research in Astronomy, Inc., under NASA contract NAS 5-03127.

The computations presented here were conducted in the Resnick High Performance Computing Center, a facility supported by Resnick Sustainability Institute at the California Institute of Technology.

The JWST data presented in this paper were obtained from MAST at STScI. The specific observations analyzed can be accessed via \dataset[DOI: 10.17909/gark-9x94]{https://doi.org/10.17909/gark-9x94}. Support for MAST for non-HST data is provided by the NASA Office of Space Science via grant NNX09AF08G and by other grants and contracts.

Resources supporting this work were provided by the (i) NASA Scientific Computing project through the NASA Center for Climate Simulation (NCCS) at Goddard Space Flight Center and (ii) Carnegie Science Earth and Planets Laboratory. This content is solely the responsibility of the authors and does not necessarily represent the views of the NCCS or Carnegie Science.

CIC acknowledges support by NASA Headquarters through (i) an appointment to the NASA Postdoctoral Program at the Goddard Space Flight Center, administered by ORAU through a contract with NASA and (ii) under award number 80GSFC24M0006. Goddard affiliates acknowledge support from the GSFC Sellers Exoplanet Environments Collaboration (SEEC), which is supported by NASA's Planetary, Astrophysics and Heliophysics Science Divisions’ Research Program.

\end{acknowledgments}

\begin{contribution}

S.K. lead this work and the Bayesian retrievals. C.I.C. performed the spectral reductions, while J.L.Y. helped interpret the retrieval results. N.L.W. helped estimate the flux calibrated stellar spectra, and G.G.C. helped perform the forward modelling and grid retrievals. S.M. performed the bulk metallicity estimation in consultation with R.H. All the co-authors assisted in interpretation of the data analysis, critically evaluating conclusions, and suggesting revisions that helped clarify the draft while improving its readability.


\end{contribution}

%

\facilities{JWST}

\software{
\texttt{astroquery} \citep{ginsburg_astroquery_2019}, 
\texttt{astropy} \citep{robitaille_astropy:_2013, astropy_collaboration_astropy_2018},
\texttt{batman} \citep{kreidberg_batman_2015},
\texttt{dynesty} \citep{speagle_dynesty_2020},
\texttt{Eureka!} \citep{bell_eureka_2022},
\texttt{EXOFASTv2} \citep{eastman_exofastv2_2019},
\texttt{ExoTiC-JEDI} \citep{alderson_early_2022, alderson_exo-ticexotic-jedi_2022},
\texttt{GASTLI} \citep{acuna_gastli_2024},
\texttt{matplotlib} \citep{hunter_matplotlib:_2007},
\texttt{MultiNest} \citep{feroz_multinest_2009},
\texttt{numpy} \citep{oliphant_numpy:_2006},
\texttt{pandas} \citep{mckinney-proc-scipy-2010},
\texttt{POSEIDON v1.2.1} \citep{macdonald_hd_2017, macdonald_poseidon_2023},
\texttt{PyMultiNest v2.12} \citep{buchner_x-ray_2014},
\texttt{scipy v1.15.2} \citep{oliphant_python_2007, virtanen_scipy_2020},
}


\appendix

\setcounter{figure}{0}
\renewcommand{\thefigure}{A\arabic{figure}}
\setcounter{table}{0}
\renewcommand{\thetable}{A\arabic{table}}

\section{Transit Fit Results}

In this section, we compile results for our transit fits that do not include any heterogeneities \autoref{apptab:NoSpotTransitfit}. Subsequently, we compare the goodness-of-fits for different number of heterogeneities used while modelling the transits for the two visits in \autoref{apptab:spotevidence}, and list the properties of the best-fitting heterogeneity model in \autoref{apptab:spotpar}.

\startlongtable
\begin{deluxetable*}{lcccc}
\tabletypesize{\footnotesize}
\tablecaption{Orbital elements based on a fit to the white light curves assuming a circular orbit and a uniform stellar surface, i.e., no heterogeneities. The fit is shown in \autoref{fig:NoSpotTransitFit}. \label{apptab:NoSpotTransitfit}}
\tablehead{\colhead{Parameter} &
\colhead{Units} &
\colhead{Prior} & 
\colhead{Visit 1} & 
\colhead{Visit 2}
}
\startdata
~~~Period & days & Fixed & \multicolumn2c{2.930289} \\
~~~Time of mid-transit ($T_0$) & $\mathrm{BJD_{TDB}}$ & $\mathcal{N}(2460489.17,0.01)$ & $2460488.433292 \pm 0.000004$ & $2460488.433406 \pm 0.000004$ \\
~~~Scaled semi-major axis ($a/R_\star$) & \nodata & $\mathcal{N}(14.1,2.0)$ & $14.12_{-0.03}^{+0.02}$ & $14.21_{-0.04}^{+0.01}$ \\
~~~Scaled radius ($R_p/R_\star$) & \nodata & $\mathcal{U}(0,1)$ & $0.2214_{-0.0011}^{+0.0006}$ & $0.2237_{-0.0012}^{+0.0004}$\\
~~~Impact parameter ($b$) & \nodata & $\mathcal{U}(0,1)$ & $0.13_{-0.01}^{+0.02}$ & $0.05_{-0.03}^{+0.04}$\\
~~~Inclination ($i$)$^b$ & deg & \nodata & $89.47_{-0.06}^{+0.04}$ & $89.8_{-0.2}^{+0.1}$\\
~~~Linear limb darkening coefficient ($q_1$)$^c$ & \nodata & $\mathcal{N}(0.115,0.1)$ & $0.115 \pm 0.005$ & $0.094_{-0.004}^{+0.005}$\\
~~~Quadratic limb darkening coefficient ($q_2$)$^c$ & \nodata & $\mathcal{N}(0.136,0.1)$ & $0.20 \pm 0.01$ & $0.37 \pm 0.02$ \\
\enddata
\tablenotetext{b}{With \texttt{juliet}, inclination is a derived parameter and thus has no prior.}
\tablenotetext{c}{Quadratic limb darkening coefficients for all white light curve fits using \texttt{juliet} were sampled with the $q_1$ and $q_2$ parameterization from \cite{kipping_efficient_2013} centered on the values predicted with \texttt{ExoTiC-LD}.}
\end{deluxetable*}

\startlongtable
\begin{deluxetable*}{cccccc}
\tablecaption{Statistical metrics for various surface heterogeneity configurations\label{apptab:spotevidence}}
\tablehead{\colhead{Number of surface features} &
\colhead{$\ln Z$} &
\colhead{$\ln B$} &
\colhead{$\mathrm{BIC}$} &
\colhead{$\mathrm{AIC}$} &
\colhead{RMSE (ppm)}
}
\startdata
\sidehead{Visit 1 (Obs 11):}
~~~0 & 18420.8 & 365.0 & $-36884.8$ & $-36944.3$ & 335.6\\
~~~1 & 18563.4 & 222.4 & $-37156.9$ & $-37240.2$ & 324.3\\
~~~2 & 18731.6 & 54.2  & $-37528.5$ & $-37635.6$ & 309.1\\
~~~\textbf{3} & \textbf{18785.8} & \nodata   & $\mathbf{-37615.0}$ & $\mathbf{-37745.9}$ & \textbf{304.5}\\
~~~4 & 18761.3 & 24.5  & $-37552.9$ & $-37707.6$ & 305.7\\
~~~5 & 18757.4 & 28.4  & $-37577.1$ & $-37755.6$ & 303.5\\
~~~6 & 18743.8 & 42.1  & $-37431.1$ & $-37633.5$ & 307.9\\
\sidehead{Visit 2 (Obs 20):}
~~~0 & 18689.6 & 98.1  & $-37430.5$ & $-37490.0$ & 315.1\\
~~~1 & 18777.3 & 10.4  & $-37597.6$ & $-37680.9$ & 307.5\\
~~~\textbf{2} & \textbf{18787.7} & \nodata  & $\mathbf{-37623.6}$ & $\mathbf{-37730.7}$ & \textbf{305.2}\\
~~~3 & 18800.2 & -12.4 & -37603.9 & -37734.8 & 304.8\\
~~~4 & 18768.5 & 19.2  & -37548.9 & -37703.6 & 305.6\\
~~~5 & 18786.1 & 1.7   & -37543.8 & -37722.3 & 304.7 \\
~~~6 & 18791.4 & -3.7  & -37518.9 & -37721.2 & 304.3\\
\enddata
\tablecomments{The adopted configuration is determined using the BIC and in bold. The Bayes factor was calculated as $\ln B=\ln Z_{\mathrm{bold}} - \ln Z_i$, where $i$ refers to the number of surface heterogeneities. RMSE is the root mean square error of the residuals.}
\end{deluxetable*}

\startlongtable
\begin{deluxetable*}{lhcccccc}
\tablecaption{Best-fitting surface heterogeneities for TOI-5293A. Based on the flux ratios, Visit 1 has two faculae ($f_{\mathrm{het},i}$>1) and one spot ($f_{\mathrm{het},i}$<1). Visit 2 has one spot and facula.\label{apptab:spotpar}}
\tablehead{\colhead{Name} &
\nocolhead{Units} &
\colhead{Prior} &
\colhead{Value} &
}
\startdata
\sidehead{Visit 1 (Obs. 11) parameters:}
~~~$f_\mathrm{het,1}$ & \nodata & $\mathcal{U}(0,2)$ & $0.2 \pm 0.1$\\
~~~$X_{1}$ & $R_\star$ & Unit Disk & $0.535 \pm 0.008$\\
~~~$Y_{1}$ & $R_\star$ & Unit Disk & $-0.41 \pm 0.02$\\
~~~$r_{\mathrm{het},1}$ & $R_\star$ & $\mathcal{U}(0,0.5)$ & $0.04 \pm 0.01$\\\hline
~~~$f_\mathrm{het,2}$ & \nodata & $\mathcal{U}(0,2)$ & $1.16_{-0.02}^{+0.03}$\\
~~~$X_{2}$ & $R_\star$ & Unit Disk & $0.189 \pm 0.007$\\
~~~$Y_{2}$ & $R_\star$ & Unit Disk & $-0.48_{-0.03}^{+0.04}$\\
~~~$r_{\mathrm{het},2}$ & $R_\star$ & $\mathcal{U}(0,0.5)$ & $0.16_{-0.04}^{+0.03}$\\\hline
~~~$f_\mathrm{het,3}$ & \nodata & $\mathcal{U}(0,2)$ & $1.14 \pm 0.021$\\
~~~$X_{3}$ & $R_\star$ & Unit Disk & $-0.905_{-0.009}^{+0.011}$\\
~~~$Y_{3}$ & $R_\star$ & Unit Disk & $-0.35_{-0.04}^{+0.03}$\\
~~~$r_{\mathrm{het},3}$ & $R_\star$ & $\mathcal{U}(0,0.5)$ & $0.30_{-0.03}^{+0.04}$\\
\hline \hline
\sidehead{Visit 2 (Obs. 20) parameters:}
~~~$f_\mathrm{het,1}$ & \nodata & $\mathcal{U}(0,2)$ & $0.95 \pm 0.01$\\
~~~$X_{1}$ & $R_\star$ & Unit Disk & $-0.50 \pm 0.02$\\
~~~$Y_{1}$ & $R_\star$ & Unit Disk & $-0.41 \pm 0.04$\\
~~~$r_{\mathrm{het},1}$ & $R_\star$ & $\mathcal{U}(0,0.5)$ & $0.23 \pm 0.04$\\\hline
~~~$f_\mathrm{het,2}$ & \nodata & $\mathcal{U}(0,2)$ & $1.12_{-0.05}^{+0.06}$\\
~~~$X_{2}$ & $R_\star$ & Unit Disk & $0.15_{-0.01}^{+0.02}$\\
~~~$Y_{2}$ & $R_\star$ & Unit Disk & $-0.19 \pm 0.03$\\
~~~$r_{\mathrm{het},2}$ & $R_\star$ & $\mathcal{U}(0,0.5)$ & $0.046_{-0.009}^{+0.012}$\\
\enddata
\tablecomments{All heterogeneities are indicated by an index, $i$, such that the corresponding flux ratio is $f_{\mathrm{het},i}$, position is ($X_i$,$Y_i$), and radius is $r_{\mathrm{het},i}$. No heterogeneities share parameters. For all surface heterogeneities, the center is sampled from a unit disk with dimensionless parameters $U_i$ and $V_i$ that were sampled with uniform priors. The positional inputs to \texttt{spotrod} are calculated as $X_i=\sqrt{U_i}\cos(2\pi V_i)$ and $Y_i=\sqrt{U_i}\sin(2\pi V_i)$. For ease of reference, we only report the parameters that would be input to \texttt{spotrod} when generating a model.}
\end{deluxetable*}

\section{Ancillary spectroscopic parameters} \label{app:ldspot}
\begin{figure}[tt]
\epsscale{0.85}
\plotone{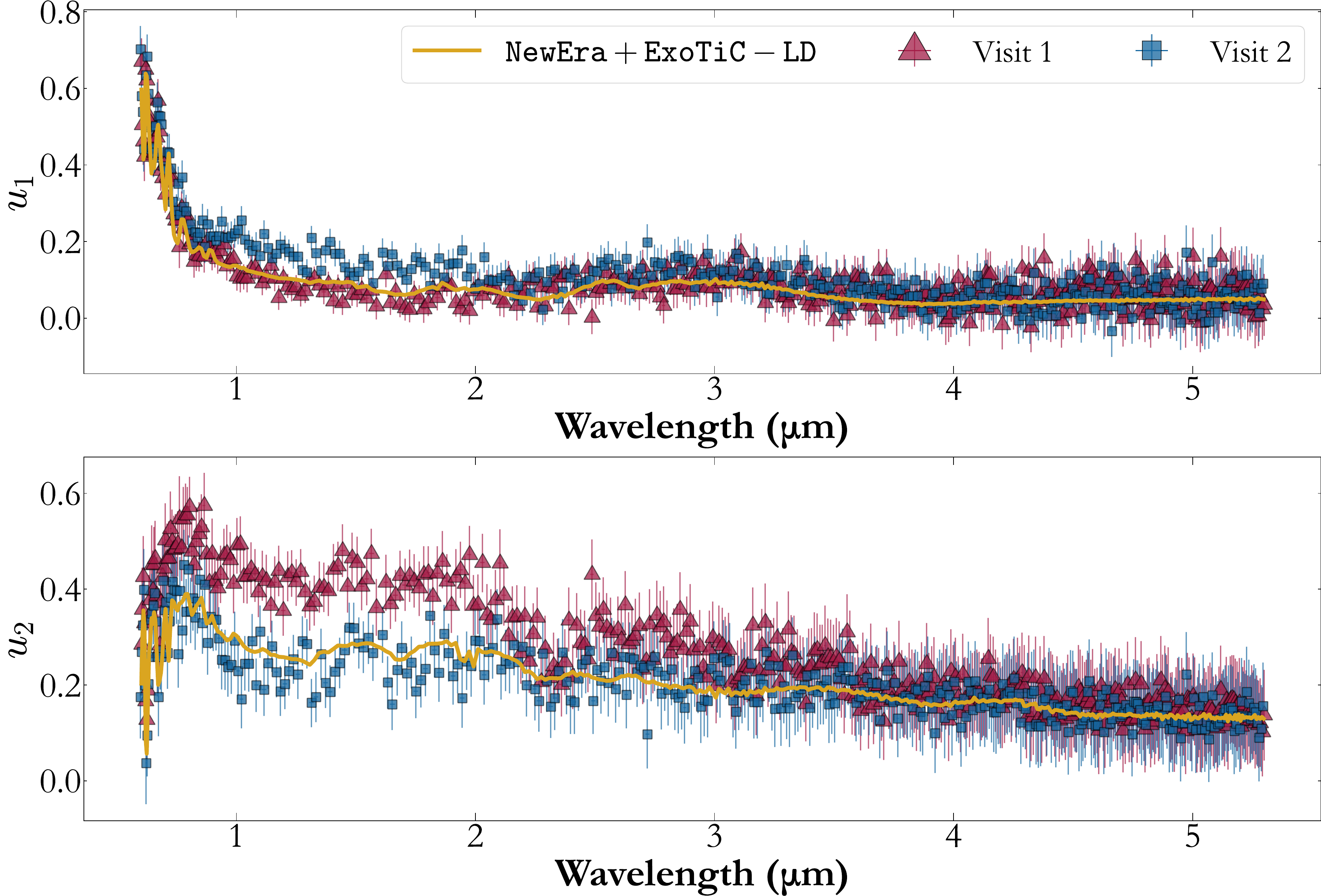}
\caption{Quadratic limb darkening coefficients,  \textbf{(top)} $u_1$ and \textbf{(bottom)} $u_2$, for the pixel-level \texttt{ExoTiC-JEDI} reductions. The derived coefficients for visit 1 (red triangles) and visit 2 (blue squares) and prediction using \texttt{ExoTiC-LD} (solid line).}
\label{fig:ldspec}
\end{figure}

\autoref{fig:ldspec} presents a comparison between the limb-darkening coefficients for a quadratic limb darkening law derived with \texttt{ExoTiC-LD} using \texttt{NewEra} models along with the retrieved values from the \texttt{ExoTiC-JEDI} reduction. The most significant differences between the retrieved values from the prediction occur at wavelengths $<2$ \textmu{}m.

\begin{figure}[tt]
\epsscale{0.85}
\plotone{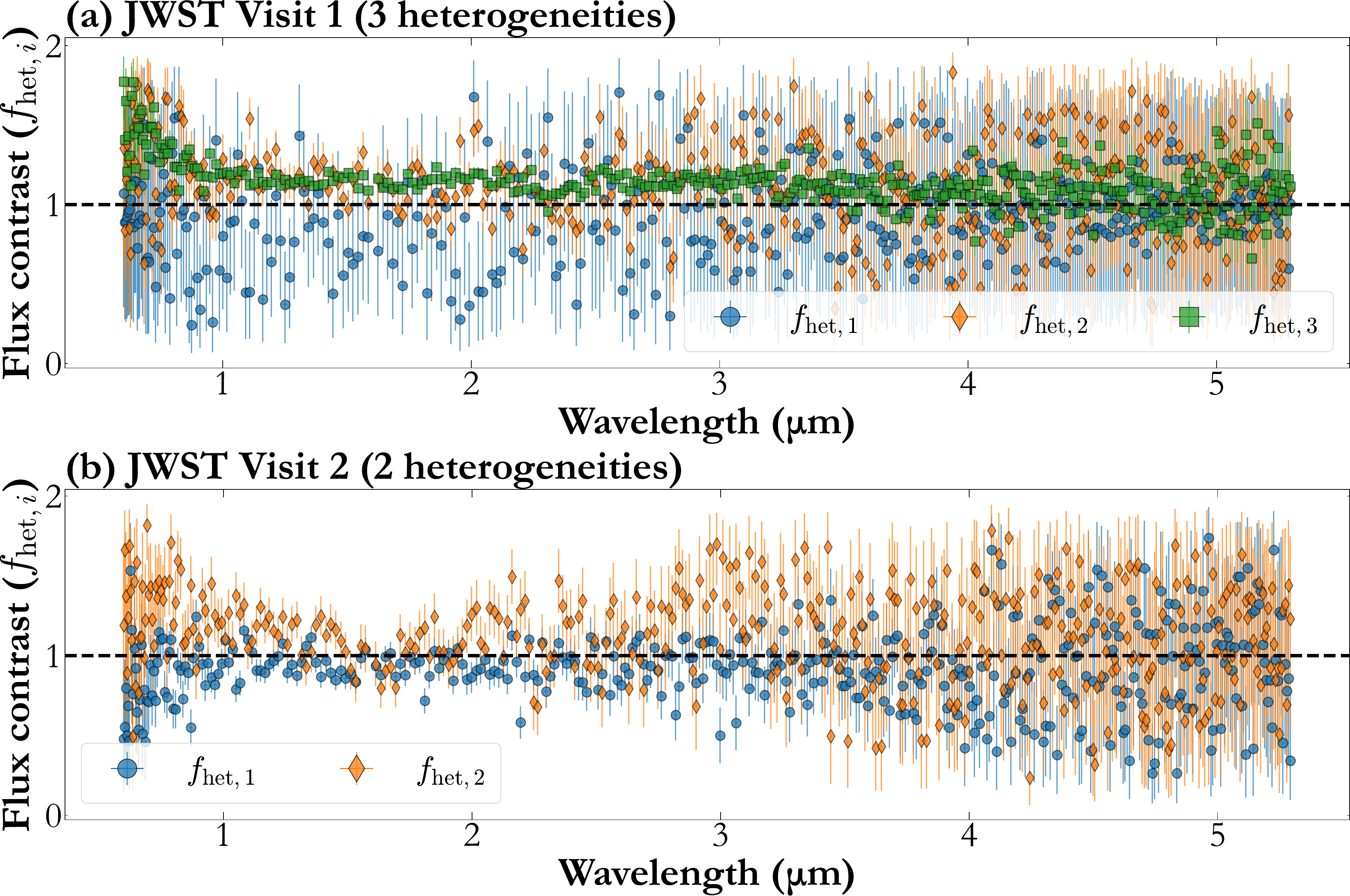}
\caption{Flux ratios for surface heterogeneities derived using the pixel-level \texttt{ExoTiC-JEDI} reductions for \textbf{(a)} Visit 1 and \textbf{(b)} Visit 2. Note that the location of heterogeneity 1  on the limb for Visit 1 (\autoref{fig:transitcrossing}) complicates the estimation of the flux ratio by introducing covariances with the limb darkening parameters (despite using the \texttt{NewEra} stellar library as priors; \autoref{fig:ldspec}).}
\label{fig:fluxconspec}
\end{figure}

\autoref{fig:fluxconspec} displays the flux ratios for all heterogeneities for each visit showing increasing (decreasing) slopes at bluer wavelengths due to spots (faculae). 

\section{Retrieval Results}

\begin{figure*}
    \centering
    \includegraphics[width=0.99\linewidth]{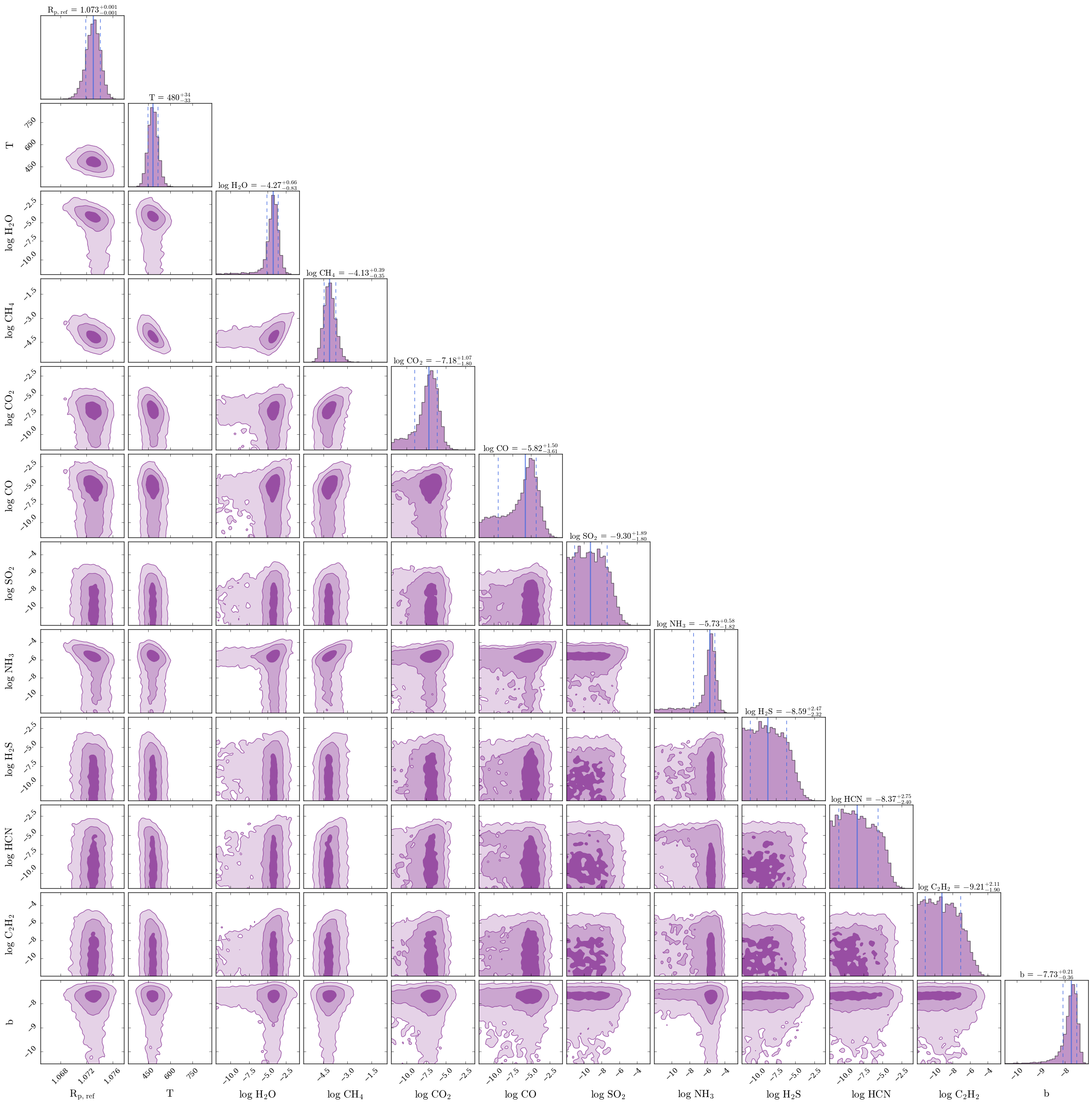}
    \caption{Full posterior probability distributions for the planetary and fitting parameters for Model 2.1 for Visit 2, which includes a planetary atmosphere without clouds or TLS components.}
    \label{fig:CornerPlotVisit2_M2.1}
\end{figure*}

\begin{figure}
    \centering
    \includegraphics[width=0.8\textwidth]{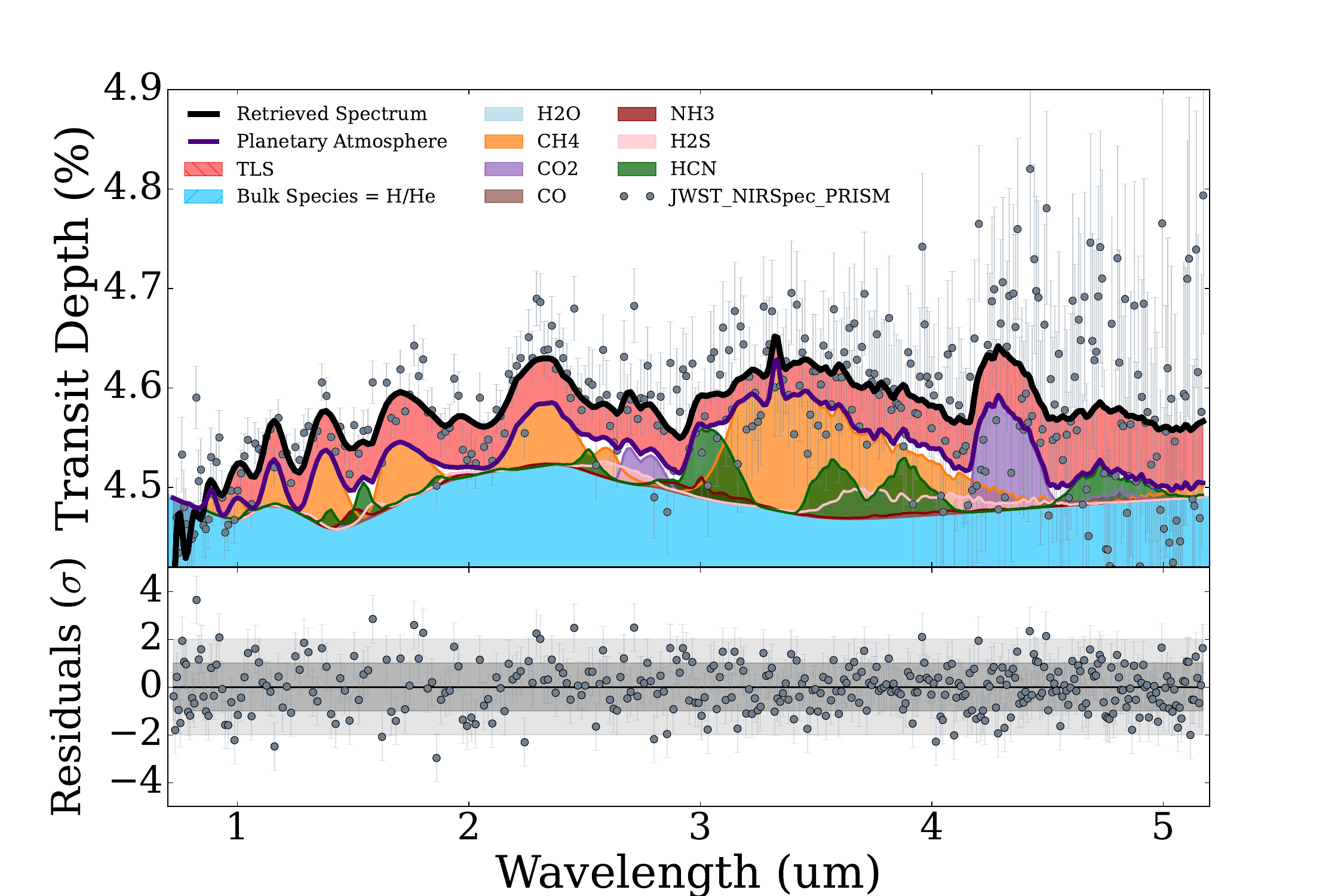}
    \hfill
    \includegraphics[width=0.5\textwidth]{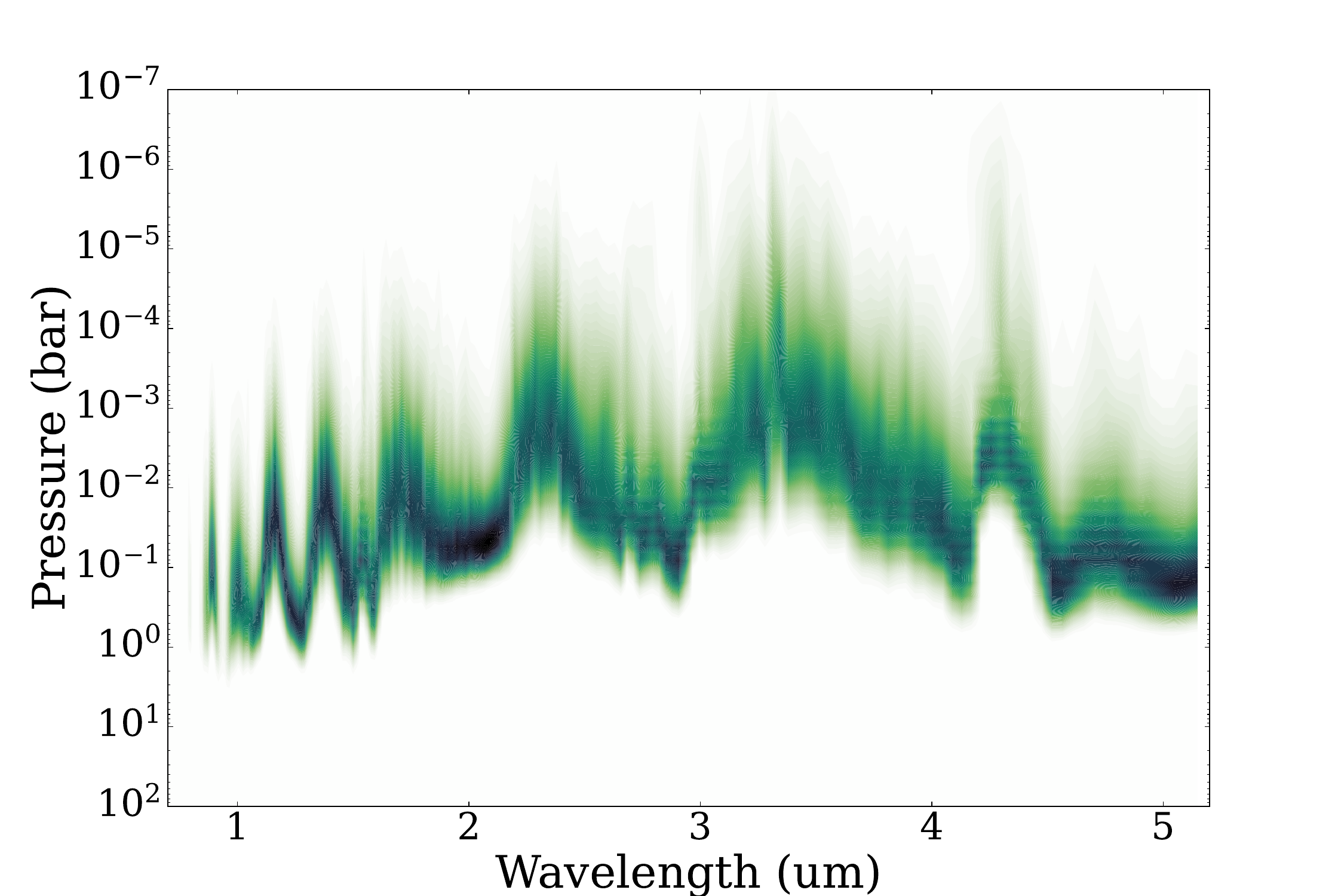}
    \caption{Retrieval results from the preferred models for Visit 1. Plot formatting and content is similar to \autoref{fig:retrieval_contributions_Visit2}. 
   \textbf{Takeaway:} \textbf{We see a difference in the recovered atmospheric properties between Visits 1 and 2 for TOI-5293Ab, which is likely due to the temporal separation between transits $\sim$ half the stellar rotation period resulting in the planet transiting across very different stellar surface along the transit chords. While we perform these retrievals on Visit 1, these should be interpreted with caution.}}
    \label{appfig:retrieval_contributions_Visit1}
\end{figure}

\begin{figure*}
    \centering
    \includegraphics[width=0.99\linewidth]{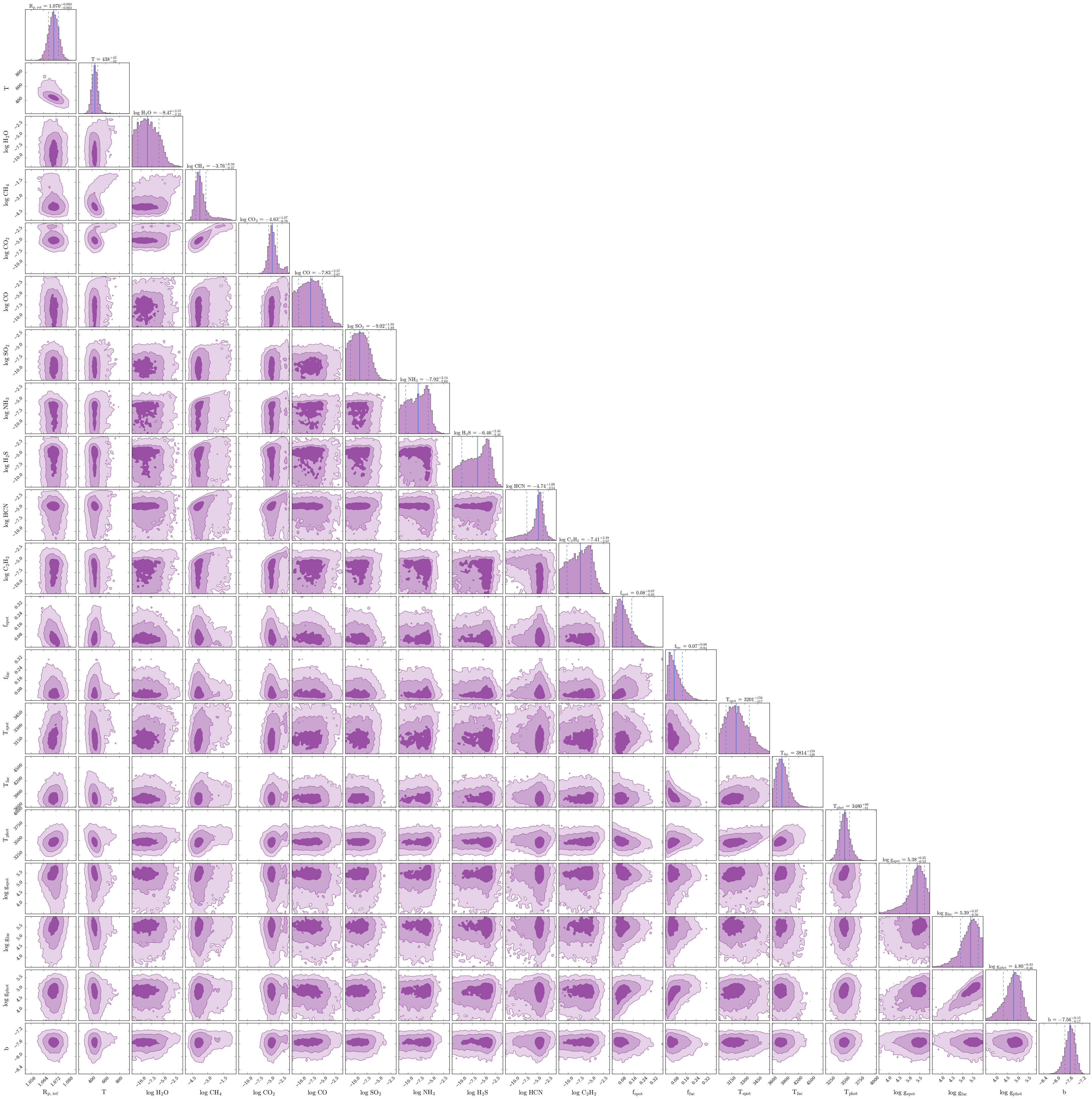}
    \caption{Similar to \autoref{fig:CornerPlotVisit2_M2.1}, but for Model 3.1 for Visit 1, with a TLS component, and a clear planetary atmosphere, i.e., no clouds.}
    \label{fig:CornerPlotVisit1_M3.1}
\end{figure*}

\begin{figure*}
    \centering
    \includegraphics[width=0.99\linewidth]{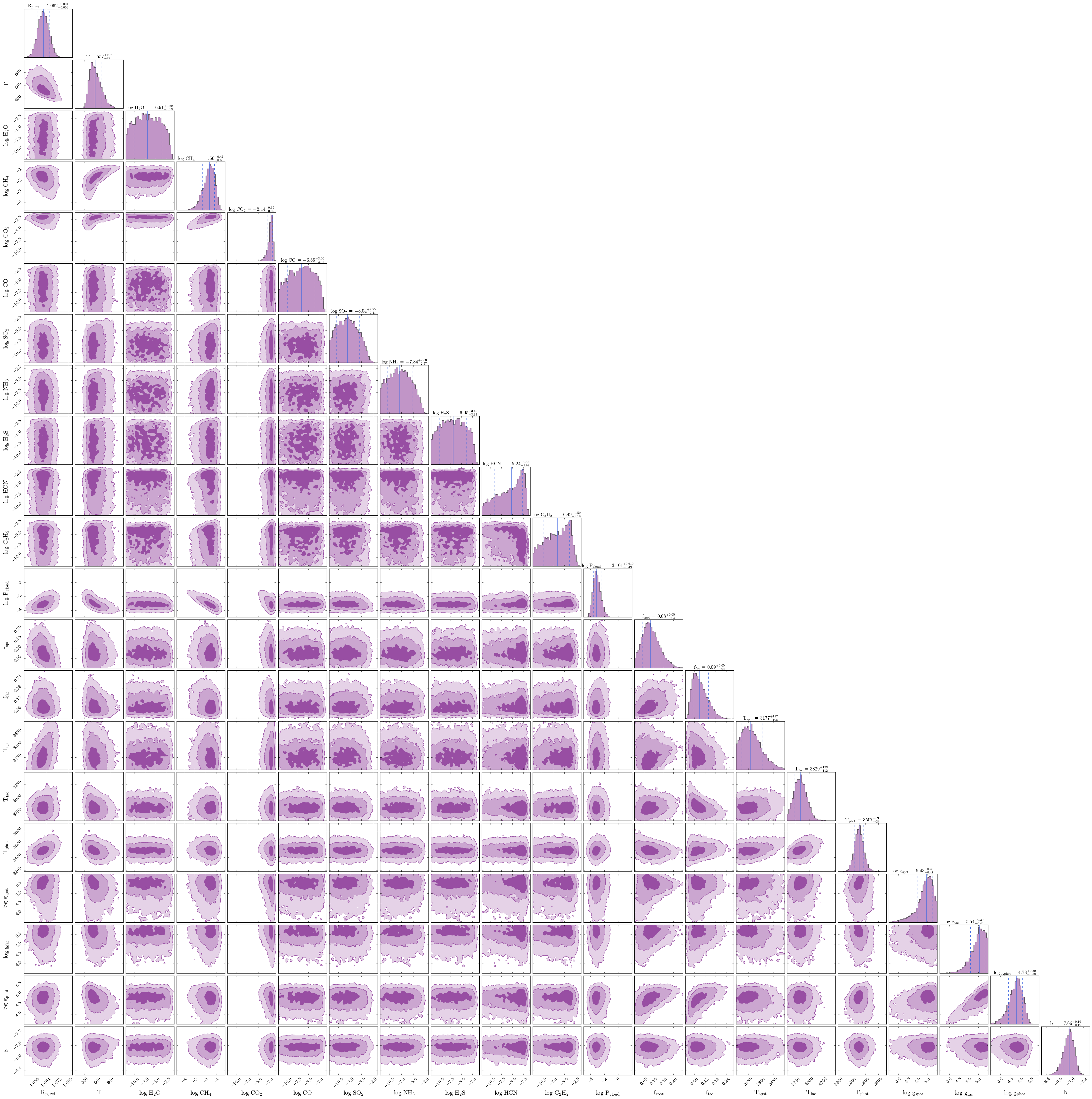}
    \caption{For comparison to the previous figure, we also include posteriors for Model 3.2 for Visit 1, with a TLS component, and planetary atmosphere including clouds.}
    \label{fig:CornerPlotVisit1_M3.2}
\end{figure*}

\bibliography{MyLibrary,morerefs}
\bibliographystyle{aasjournalv7}



\end{document}